\documentclass[aps,floatfix,superscriptaddress,twocolumn,amsmath,amssymb,nofootinbib,10pt,longbibliography]{revtex4-1}

\usepackage{enumitem}
\usepackage{multirow} 
\usepackage{float}
\usepackage{setspace}
\usepackage[dvipsnames,svgnames]{xcolor}
\usepackage[normalem]{ulem}
\usepackage{silence}
\usepackage{hyperref}
\definecolor{refColour}{rgb}{0.4, 0.19, 0.28}
\definecolor{citeColour}{rgb}{0.5,0.5,0.0}
\definecolor{citeColour}{rgb}{0.41, 0.16, 0.38}
\definecolor{extraColour}{rgb}{0, 0, 0}
\hypersetup{colorlinks=true,linkcolor=blue,citecolor=blue}

\def\beq{\begin{equation}}
\def\eeq{\end{equation}}
\def\barray{\begin{eqnarray}}
\def\earray{\end{eqnarray}}

\usepackage{csquotes}
\usepackage{natbib}
\usepackage[caption=false]{subfig}
\usepackage{epigraph}
\usepackage{fontawesome}
\usepackage{graphicx}
\usepackage{ulem}
\usepackage{quantikz}
\usepackage{comment}

\usepackage{siunitx}
\usepackage[caption=false]{subfig}
\begin{document}

\title{Photonic Implementation of the Quantum Morra Game}

\author{Andrés Ulibarrena}
\affiliation{Institute of Photonics and Quantum Sciences, School of Engineering and Physical Sciences, Heriot-Watt University, Edinburgh EH14 4AS, United Kingdom}

\author{Alejandro Sopena}
\affiliation{Instituto de Física Teorica, UAM-CSIC, Universidad Autónoma de Madrid, Cantoblanco, 28049 Madrid, Spain}

\author{Russell Brooks}
\affiliation{Institute of Photonics and Quantum Sciences, School of Engineering and Physical Sciences, Heriot-Watt University, Edinburgh EH14 4AS, United Kingdom}

\author{Daniel Centeno}
\affiliation{Instituto de Física Teorica, UAM-CSIC, Universidad Autónoma de Madrid, Cantoblanco, 28049 Madrid, Spain}
\affiliation{Perimeter Institute for Theoretical Physics, Waterloo, Ontario, Canada, N2L 2Y5}
\affiliation{Department of Physics and Astronomy, University of Waterloo, Waterloo, Ontario, Canada, N2L 3G1}

\author{Joseph Ho}
\affiliation{Institute of Photonics and Quantum Sciences, School of Engineering and Physical Sciences, Heriot-Watt University, Edinburgh EH14 4AS, United Kingdom}

\author{Germ\'{a}n Sierra}
\affiliation{Instituto de Física Teorica, UAM-CSIC, Universidad Autónoma de Madrid, Cantoblanco, 28049 Madrid, Spain}

\author{Alessandro Fedrizzi}
\affiliation{Institute of Photonics and Quantum Sciences, School of Engineering and Physical Sciences, Heriot-Watt University, Edinburgh EH14 4AS, United Kingdom}

\begin{abstract}
    The Morra game, an age-old non-cooperative game, traditionally played on one's hand, has proved to be a rich setting to study game-theoretic strategies, both classically and within the quantum realm.
    In this paper, we study a faithful translation of a two-player quantum Morra game, which builds on previous work by including the classical game as a special case. We propose a natural deformation of the game in the quantum regime in which Alice has a winning advantage, breaking the balance of the classical game. A Nash equilibrium can be found in some cases by employing a pure strategy, which is impossible in the classical game where a mixed strategy is always required. We prepared our states using photonic qubits on a linear optics setup, with an average deviation $\leq 2 \%$ with respect to the measured outcome probabilities. Finally, we discuss potential applications of the quantum Morra game to the study of quantum information and communication. 
\end{abstract}

\maketitle

\section{\label{sec:introduction}Introduction}
Modern game theory, initially developed by J. Nash and J. Neumann~\cite{von1947theory,nash} for analysing economic strategies among rational players in a mathematically consistent model, has found widespread applications in various fields, including biology~\cite{ghirlanda2004evolution}, politics~\cite{politics}, choice-theory~\cite{choice} and computer science~\cite{shoham2008computer}. 
However, the emergence of quantum information in the 80's and 90's enabled researchers to study game theoretic models in the quantum regime and it has turned out that superposition and entanglement allow new winning strategies that previously did not exist~\cite{Meyer,INGARDEN197643}.
Quantum games (QG) have already been used to design quantum cryptographic protocols with enhanced security~\cite{Bennett2014, Wiesner} and optimise networks for frequency allocation~\cite{zabaleta2017quantum}.
Moreover, studying QG's has revealed new insights on non-cooperative games such as prisoners dilemma where a new equilibrium strategy was found and improved the payoff to all players~\cite{Eisert,benjamin2001multiplayer}. Quantum games can now be implemented on plethora of systems, such as small-scale quantum processors~\cite{anand2020prisonerIBM}, ion traps~\cite{solmeyer2018demonstration} or nitrogen vacancies in diamonds~\cite{george2013threeboxes} however the most common platform has been photonics~\cite{zeilinger2007prisoner,pinheiro2013vectorvortex,schmid2010experimentalminority,balthazar_experimental_2015}. This has the benefit that the games can be played between remote  players, and that they can then be used as primitives for higher level distributed quantum communication tasks~\cite{dey2023quantum}.  

Here we consider one of the oldest known games that is still studied today, the Morra game~\cite{suetonius2019lives,morris2012introduction}. Morra is a non-cooperative game in which players hide a maximum number of coins (\textit{or fingers}), and each player attempts to guess the total number. 
Players are ordered a priori, and the rule is that a player cannot repeat the guess of the previous ones.
This rule apparently gives the first player an advantage over the rest, but unexpectedly everyone has an equal chance of winning. The Morra is a closely related game to the Spanish Chinos game, which has been studied extensively~\cite{Pastor, pastor2003learning}.
The Morra's game has applications in modelling financial markets, and information transmission~\cite{Pastor}. 
 
Quantum versions of Morra's game have been proposed in the past~\cite{GMD,centeno2022Chinos}, with the aim of studying the equilibrium strategies between the two players.
The basic idea is to replace the act of drawing a coin by that of applying an operator on a quantum state shared by all players, say a boson, or a hard-core boson (fermions), a qubit, or two qubits. These operations create a state in which one measures an observable that is the classical analogue of the total number of coins. In some situations quantum effects may lead to the breaking of the classical balance of the players.

Here we show a novel implementation of the quantum Morra game (QMG) using qutrits that, unlike the previous proposals~\cite{GMD,centeno2022Chinos}, can reproduce the classical game faithfully. This version of the game allows us to deform the underlying rules, which we shall define later, thereby generating new effects. This implementation makes use of a three level system as shared resource for players to act on, similar to the Aharonov three quantum boxes game~\cite{george2013threeboxes}. We implement this game employing entangled photonic qubits and obtained good agreement between the theory and the experimental results.

\subsection{\label{sec:CMG}Classical Morra Game}
We shall consider the simplest version of the Morra game that involves two players, Alice and Bob, who can draw each from 0 to 1 coins and guess the total number of coins with the restriction that Bob cannot repeat the result predicted by Alice. This game can be generalised to more players and coins~\cite{GMD}. We distinguish between pure strategies, which involve players selecting a specific number of coins, and mixed strategies, where players randomize their coin choices based on probabilities assigned to each possible number of coins.
A non-cooperative game which allows mixed strategies is guaranteed to have at least one Nash equilibrium~\cite{glicksberg1952further}, a situation where neither player can improve their payoff by unilaterally changing their strategy, as long as their opponent does not change their strategy.
We refer to a strategy as optimal when it leads to a Nash equilibrium~\cite{nash}.

The optimal strategy for Alice is to choose at random $c_A=0,1$ coins and to guess always $g_{A} =1$, so as not to reveal information to Bob~\cite{GMD}. This is based on the fact that with four possible outcomes, the most probable value of the sum is 1. Bob's  optimal strategy is to choose  $c_B=0,1$ coins at random and make his guess $g_{B}$ in a rational way~\cite{mele_oxford_2004}.

A rational player seeks to maximise their expected payoff or benefit~\cite{askari2019behavioral}. For example, users competing for bandwidth in a radio network~\cite{naseer2021game}, or prisoners minimising their jail sentence~\cite{Eisert}. In the context of the Morra game, if Bob chooses $c_B=0$, then by reason he must exclude the option $g_{B}=2$, and if he chooses $c_B=1$, then he must exclude $g_B=0$. Playing the optimal strategy, each player wins on average half of the time, resulting in a symmetric game with equal winning probabilities. 

This strategy is also Pareto optimal~\cite{pareto}, since no player can improve their own payoff without reducing their opponents payoff. In game theory, this scenario is called a zero-sum game~\cite{von1947theory}. This result is also valid for two players and a general number of coins~\cite{GMD}.

\section{\label{sec:QMG}Quantum Morra Game}
In our version of the QMG, first we will associate the total number of coins, namely 0, 1, 2, to three orthogonal states
$|\tilde{0}\rangle, |\tilde{1} \rangle, | \tilde{2} \rangle$ that form the basis of a qutrit.
To produce these states, the players have two unitary operators, ${\cal O}_0$ and ${\cal O}_1$,
that correspond to the number of coins they have in their hands at the start of each roll.
The joint state created by Alice and Bob is given by
\beq
|\psi_{a,b} \rangle = {\cal O}_a^A {\cal O}_b^B |\phi \rangle, \qquad a, b =0,1  \, , 
\label{t4}
\eeq  
where $|\phi \rangle$ is an initial state and note $|\psi_{0,1} \rangle$ and $|\psi_{1,0} \rangle$ are identical states (a relative phase between them is unobservable).
Using this equation the basis states are obtained as
$|\tilde{0}\rangle=|\psi_{0,0}\rangle$, $|\tilde{1}\rangle=|\psi_{0,1}\rangle=|\psi_{1,0}\rangle$ and $|\tilde{2}\rangle=|\psi_{1,1}\rangle$.
Choosing ${\cal O}_0$ as the identity operator we find that $|\phi \rangle = |\tilde{0} \rangle$ and 
\beq
| \tilde{1}  \rangle = {\cal O}_1  | \tilde{0} \rangle, \quad 
| \tilde{2}  \rangle = {\cal O}_1^2  | \tilde{0} \rangle  \, . 
\label{t8}
\eeq
The operator ${\cal O}_1$ has to be unitary. A solution compatible with~\eqref{t8} is ${\cal O}_1 = X$, where $X$ is the Pauli matrix for qutrits,
\beq
X = \left( 
\begin{array}{ccc}
0 & 0 & 1 \\
1 & 0  & 0 \\
0 & 1 & 0  \\
\end{array} \right)  \, , \label{t9}
\eeq
that satisfies $X^3 = \mathbb{I}$. In the quantum game, the operator $X^3$ never arises because the maximum number of $X$ operators is 2. However, we shall consider that having three coins in the box is the same as having none.

The outcome of a round of the game is determined by measuring the observable 
\begin{equation}
    \hat{N}=\sum^{2}_{n=0}n|\tilde{n}\rangle\langle\tilde{n}| \ .
\end{equation}
The state after such measurement will be $|\tilde{0}\rangle$, $|\tilde{1}\rangle$, or $|\tilde{2}\rangle$, thereby revealing the number of coins. The winner of the round is the player whose guess matches the number of coins. Our game can be generalised to to n-players, which we show in Appendix~\ref{SM:GQMG}.
The scalability of potential moves improves within this encoding scheme compared to previous implementations~\cite{GMD,centeno2022Chinos} due to the binary choice of the coin operator. The original game employed four operators~\cite{GMD} and the single qubit game uses three operators~\cite{centeno2022Chinos}. Consequently, our game scales as $2^N$, where $N$ is the number of players, compared to the previous $4^N$ and $3^N$ for the respective games.

\subsection{\label{sec:deformation}Quantum Deformation} 
In this section we will precisely define what a deformation is in the context of our game. We are specifically referring to the players encoding operator \eqref{t9} that controls the unitary evolution of the shared state, and the underline probability distribution of the game. To go beyond the classical setting we use a parameter $\theta$ to smoothly deform the operator away from \eqref{t9}. The physical meaning of this deformation is that the players are able to toss a certain superposition of the number of coins instead of just the classical options.

We have translated the classical Morra game into a quantum game by means of a one-to-one correspondence between classical and quantum objects. 
The quantum realm allows us to define the superposition of coin states, in analogy to the  superposition of cat states. For this we use the quantum Fourier transform 
\beq
{\cal F} | \tilde{j}  \rangle = \frac{1}{ \sqrt{3}} \sum_{k=0} ^2 \omega^{ j k}   | \tilde{k}  \rangle, \quad \omega = e^{ 2 i \pi/3}  \, . 
\label{t21}
\eeq
Measuring the number of coins in each of the states~\eqref{t21} yields the values 0, 1 or 2 with equal probability $1/3$. Moreover, applying the operator $X$ to~\eqref{t21}, that is {\em adding}  a quantum coin, just multiply these states by a global  phase, $X  | \hat{j} \rangle   = \omega^{-j}  | \hat{j} \rangle$.

Along with the Pauli matrix $X$ for qutrits~\eqref{t9}, there is a
matrix $Z$ defined as $Z = \operatorname{diag}(1,  \omega,  \omega^2)$ 
that satisfies  $ZX = \omega XZ$ and is related to $X$  by the Fourier transform 
\beq
X = {\cal F} ^\dagger Z {\cal F}, 
\label{b25}
\eeq
where ${\cal F}_{j,j'} = \frac{1}{\sqrt{3}}  \omega^{j j'} \;( j,j' = 0,1,2)$.
This connection will allow us to deform the quantum game from the phase space.

The two players game can be deformed by replacing $Z$ in~\eqref{b25} by 
$Z_\theta = \textrm{diag}\left( 1, e^{ i \theta} ,  e^{ 2 i\theta} \right)$ yielding a modified $X$ operator,
\beq
X_\theta=  {\cal F} ^\dagger Z_\theta {\cal F} = 
\left( 
\begin{array}{ccc}
x_0(\theta) & x_2(\theta) & x_1(\theta) \\
x_1(\theta) & x_0(\theta) & x_2(\theta) \\
x_2(\theta) & x_1(\theta) & x_0(\theta) \\
\end{array} \right)  \, , 
\label{d2}
\eeq
where
\beq
x_j(\theta) = \frac{1}{3} (  1 + \omega^{2 j} e^{i \theta} + \omega^{ j} e^{ 2 i \theta})  \, . 
\label{d3}
\eeq
The states created by the action of $X_\theta$ are
\begin{align}
| 1_\theta \rangle &= X_\theta | \tilde{0}  \rangle = x_0(\theta) | \tilde{0}  \rangle + x_1(\theta) | \tilde{1}  \rangle +  x_2(\theta) | \tilde{2}  \rangle \, , \label{eq:d4}   \\
| 2_\theta \rangle &= X_\theta^2 | \tilde{0}   \rangle = x_0(2 \theta) | \tilde{0}  \rangle + x_1(2 \theta) | \tilde{1}  \rangle +  x_2(2 \theta) | \tilde{2}  \rangle \, , \label{eq:theta_states}
\end{align}
where we used that $X_{\theta}^2  = X_{2 \theta}$. These states are normalised but are not orthogonal except for $\theta = 2 \pi/3$ and $4 \pi/3$.
We have found a general two-qubit quantum circuit for the deformation unitary~\eqref{d2} using eight local unitaries and three CNOT gates~\cite{vidal_universal_2004,shende_recognizing_2004}, as outlined in the Appendix~\ref{SM:Decomposition}. Regarding the physical interpretation of the deformation, \eqref{eq:d4} tells us that, now, there exists the possibility of one player tossing two coins with a certain probability although in the classical version they only have one coin in their hands.

We recall that the probabilities are given by,
\beq
p_{a,b}(n)=|\langle n|\psi_{a,b}\rangle |^2
\label{d11} 
\eeq
where $|n\rangle$ are the states $|\tilde{0}\rangle, |\tilde{1} \rangle, | \tilde{2} \rangle$ and $|\psi_{a,b}\rangle$ are the common states generated by Alice and Bob using the operators $X_\theta$ and $X_\theta^2$. In particular the average probabilities of Alice are given by $P_a(n) = \frac{1}{2}\sum_b p_{a,b}(n)$, which becomes
\beq
P_a(n)=  \frac{1}{2} ( |x_n(a\theta)|^2 + |x_n((a+1)\theta)|^2 ) \
\label{d12}
\eeq

\section{\label{sec:experimental}Experimental Implementation}
We experimentally realise the QMG with a linear optics circuit, using polarisation encoded photons to prepare a two-qubit state. We can write the qutrit state $|\tilde{a} \rangle$ as a two-photon state $|i_A  i_B\rangle$ where  $(i_A , i_B = H,V)$ are the horizontal and vertical polarisation states of each photon and the indices indicate a different photon mode. The qutrit to qubit transformation can be found in the Appendix~\ref{SM:Qutrits_to_Qubits}, where we arrive at the same encoding states,
\begin{align}
| \tilde{0}  \rangle & =  |H H \rangle \, ,  \label{exp1}  \\
| \tilde{1}  \rangle & =  \frac{1}{\sqrt{2}} ( |H V \rangle + | V H \rangle ) \, , \label{exp2} \\
| \tilde{2}  \rangle & =  |V V \rangle \label{exp3}   \, ,
\end{align}

\begin{figure}   
    \centering
\includegraphics[width=0.5\textwidth]{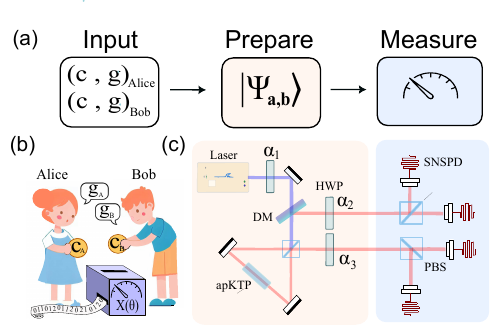}
    \caption {The QMG experimental implementation. \textcolor{blue}{(a)}  Alice and Bob's strategies are input into a device which prepares the joint state $\ket{\psi_{a,b}}$ and subsequently measures it.
    \textcolor{blue}{(b)} Alice and Bob play the QMG by secretly depositing coins $c_{A}$ and $c_{B}$ into the device which has some pre-determined parameter $\theta$. Alice makes the first guess $g_{A}$, followed by Bob $g_{B}$. The device deforms the classical coins into quantum coins using the Unitary $X(\theta)$, performs a measurement and outputs the sum of the coins. 
    \textcolor{blue}{(c)} The linear optics circuit for state preparation begins with a photon pair source, constructed with a apKTP crystal housed in a Sagnac configuration~\cite{fedrizzi2007wavelength,pickston2023Trident} pumped with a ti:sapphire laser centered at \SI{775}{\nm} and prepared in H with a linear polariser (POL), producing degenerate photon pairs at \SI{1550}{\nm}, which are seperated from the pump with a dichroic mirror (DM). The states $\ket{\psi_{a,b}}$ are encoded using the three half-wave plates (HWP) $\alpha_{1},\alpha_{2},\alpha_{3}$, outlined in the Appendix~\ref{SM:Parameters}. The measurement station consists of two polarising beam splitters (PBS) that project the two photons into the Z-basis, then coupled to superconducting nanowire single photon detectors (SNSPD), and 2-fold coincidences are identified using a time-tagging logic box.}
    \label{fig:ExpSetup}
\end{figure}

Figure~\ref{fig:ExpSetup} shows our experiment outlined in three stages; Alice and Bob input their coin choice to the device, the device prepares the shared state, and finally measures the outcome. We prepare our states~\eqref{eq:d4} and~\eqref{eq:theta_states} using the parameterised half-wave plate configuration $(\alpha_1,\alpha_2,\alpha_3)$. We used solutions that match the desired outcome probability of the shared state and preserve the entanglement entropy. We reframe the problem as an optimization problem and determine the solution with the \textit{L-BFGS-B} algorithm~\cite{byrd_limited_1995} implemented in \textsc{SciPy}~\cite{2020SciPy-NMeth}, as outlined in the Appendix~\ref{SM:Parameters}. 
These states require high fidelity and purity to recreate the game faithfully, so we take advantage of a bright photon-pair source using an aperiodically-poled potassium titanyl phosphate (apKTP) crystal, which creates spectrally pure entangled photon pairs via spontaneous parametric down-conversion (SPDC)~\cite{graffitti2018independent,Pickston:21}.

\subsection{\label{sec:results}Results} 
\begin{figure}
     \centering
     \subfloat[\label{fig:ProbsFigure1}]{
    \includegraphics[width=0.9\linewidth]{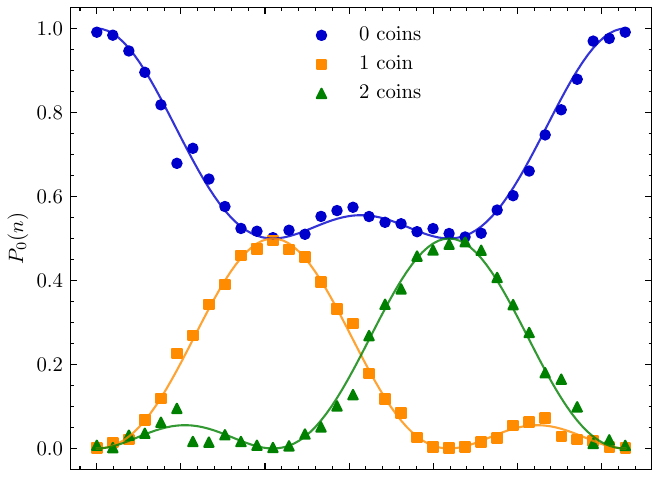}}
    \break
    \centering
    \subfloat[\label{fig:ProbsFigure2}]{ \includegraphics[width=0.9\linewidth]{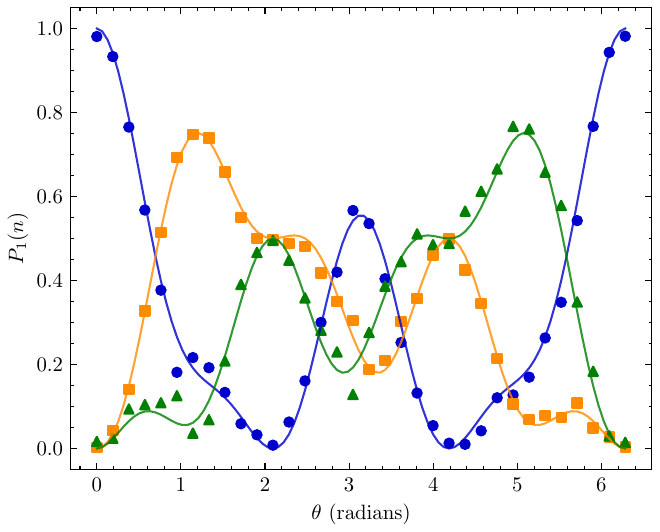}} 
 \caption{Alice's probability of winning versus $\theta$, when playing 0 and 1 coin. The probabilistic outcome of obtaining (0,1,2) coins when Alice plays no coins ($\mathbb{I}$) \protect\subref{fig:ProbsFigure1} and when she plays one coin \protect\subref{fig:ProbsFigure2} have been plotted on top of the theoretical curves. Errors bars have calculated assuming Poisson statistics, but found to be $\mathcal{O}(10^{-4})$, therefore neglected.}
  \label{fig:ProbsFigure}
\end{figure}

Approximately 150 000 rounds are played for 34 evenly spaced values of $\theta$ in the interval [0,$2\pi$]. The data collection methods are discussed in Appendix~\ref{SM:Methods}. The normalised probabilities for states $|1_\theta \rangle$ and $|2_\theta \rangle$ can be found in Appendix~\ref{SM:Parameters}. To analyse how these new outcome probabilities affect the players strategies we have averaged Bob's two plays over each of Alice's choices using~\eqref{d12}. The results have been plotted in Figure~\ref{fig:ProbsFigure1} when Alice plays zero coins, and Figure~\ref{fig:ProbsFigure2} when Alice plays one coin. The experimental values closely align with the theoretical expectation, deviating by a maximum of 2\% and  confirms a successful implementation of the QMG. We have measured a fidelity, compared to the ideal states $|\tilde{0}\rangle$, $|\tilde{1}\rangle$ and $|\tilde{2}\rangle$, of $97\%$.
The main causes for the uncertainties in our experimental data are due to factors such as imbalances on the PDC photon generation loop, losses and depolarisation on the single-mode fibres and imperfections in the retardance of the half waveplates. 

In Table~\ref{tab:ResultsTableT1} we summarise the most notable points in the Figure~\ref{fig:ProbsFigure}, where $P_i(G)$ is Alice's probability of winning by guessing $G$ when playing $i$ coins.

At $\theta$ values close to $2\pi/3$, the classical regime is recovered, matching the values predicted by the theory in Figure~\ref{fig:ProbsFigure}. 
Similarly, at $4\pi/3$, the classical probabilities are reproduced, but the guesses for coins 1 and 2 are flipped. This creates a classically impossible situation since Alice plays no coins and can expect a 50\% probability of winning by guessing 2 coins. 

Notably, for $\theta$ equalling 0 and $2\pi$, the game undergoes a complete deformation, resulting in all coin states overlapping into the 0 coin outcome, as we can see in Table~\ref{tab:ResultsTableT1}. In this case, Alice will always win if she guesses $0$ coins, regardless of the number of coins in play.

When $\theta=\pi$, the probable outcome for each coin is identical regardless of the coin Alice picks. Even so,
Alice still has an advantage over Bob as guessing $|\tilde{0}\rangle$ yields a winning result higher than 0.5.

\begin{table}[b]
\begin{ruledtabular}
  \begin{tabular}{  c  c  c  c c }      
    $\theta$  & $0$ & $2\pi/3$ & $\pi$ & $4\pi/3$ \\ 
    \colrule
    $P_0(0)$ & 0.99 & 0.50 &  0.57 & 0.51 \\  
    $P_0(1)$ & $1.19\cdot 10^{-3}$ & 0.49 & 0.30 & $1.43\cdot 10^{-3}$\\  
    $P_0(2)$ & $7.63 \cdot 10^{-3}$ & $2.46 \cdot 10^{-3}$ &  0.13 & 0.49 \\  
    $P_1(0)$ & 0.98 & $7.86 \cdot 10^{-3}$ &  0.57 & $1.26 \cdot 10^{-2}$ \\  
    $P_1(1)$ & $2.46 \cdot 10^{-3}$ & 0.50 &  0.30 & 0.49 \\  
    $P_1(2)$ & $1.67 \cdot 10^{-4}$ & 0.49 &  0.13 & 0.49 \\      
  \end{tabular}
  \caption{Alice's probability of winning ($P_i$) for different $\theta$ values, depending on the amount of coins in play ($i$).}
    \label{tab:ResultsTableT1}
\end{ruledtabular}
\end{table}

\section{\label{sec:strategies}Strategies}

The deformation $\theta$ opens up the possibility for new strategies in which Alice's winning probability is higher than 50\%.
We focus on analysing the best strategies for each player when they play randomly, as well as characterising the optimal strategy as a function of the deformation.
Each player's strategy is defined by the probability of playing zero or one coin, as well as their guesses for each situation.
Since Alice plays first, we consider strategies in which her guess is independent of the number of coins she plays, in order to not reveal information, similarly to the classical case. 
The winning probability of Alice given a strategy where she plays $a$ coins with probability $\tilde{P}^A_a$ and guesses $n$ coins, while Bob plays $b$ coins with probability $\tilde{P}^B_b$ is
\begin{equation}
    P_{win}^A = \sum_{a,b}\tilde{P}^A_a\tilde{P}^B_b p_{a,b}(n),
\end{equation}
where $p_{a,b}$ is given in~\eqref{d11}. On the contrary, Bob can change his guess $n_b$ based on the number of coins he plays, but it must be different from Alice's guess. The winning probability of Bob is
\begin{equation}
    P_{win}^B = \sum_{a}\tilde{P}^A_a(\tilde{P}^B_0 p_{a,0}(n_0)+\tilde{P}^B_1 p_{a,1}(n_1)).
\end{equation}
We say that a player's strategy is stable when his probability of winning is higher than that of the other players even if they change their strategies.

\begin{figure*}[t]
\centering
    \subfloat[\label{fig:Strategy_random}]{
        \includegraphics[width=0.34334\textwidth]{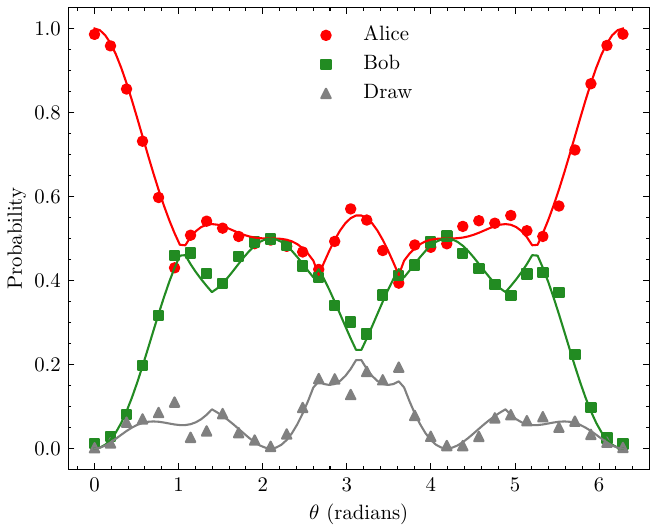}} 
    \subfloat[\label{fig:Strategy_a}]{
        \includegraphics[width=0.33\textwidth]{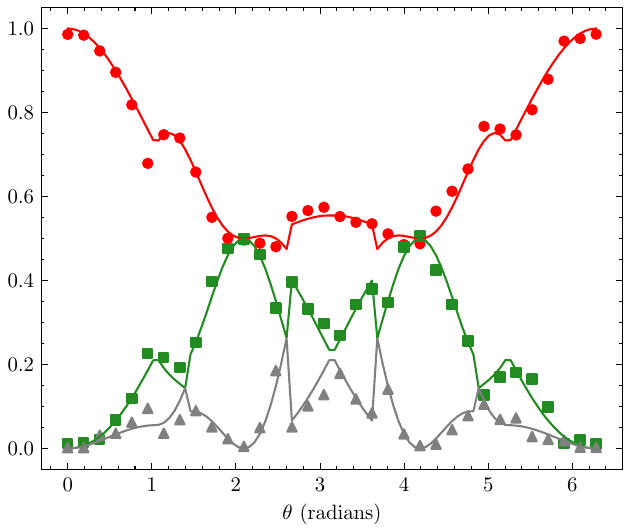}}
    \subfloat[\label{fig:Strategy_b}]{
        \includegraphics[width=0.33\textwidth]{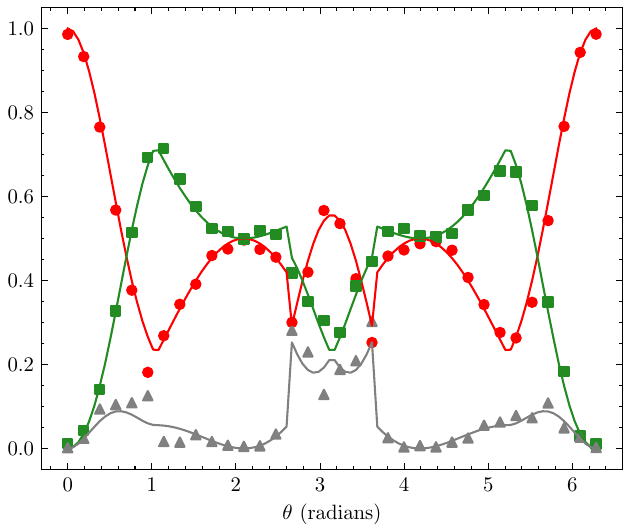}}
\caption{\label{fig:strategies}\protect\subref{fig:Strategy_random} The winning probabilities of Alice (red) and Bob (green), as well as the probability of a draw (gray), when both players choose 0 or 1 coin at random but make their best guess. \protect\subref{fig:Strategy_a} Alice plays her best strategy knowing that Bob plays randomly. \protect\subref{fig:Strategy_b} Bob plays his best strategy knowing that Alice plays randomly. The experimental results have been plotted on top of the theoretical values.}
\end{figure*}

\begin{figure}
     \centering
    \includegraphics[width=0.8\linewidth]{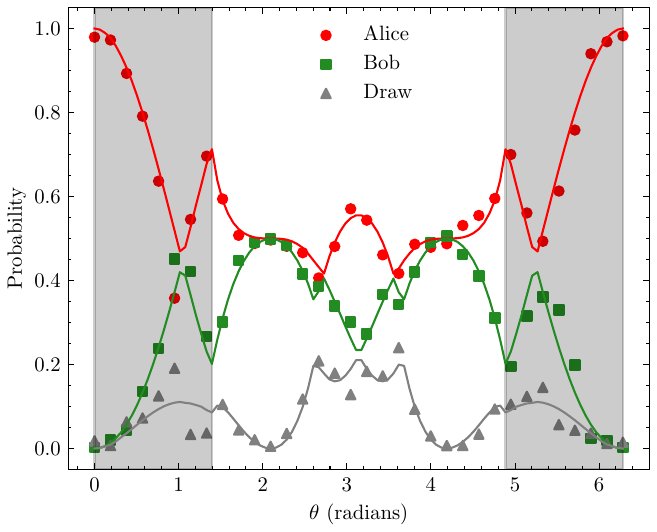}
 \caption{Optimal strategies for Alice and Bob. The winning probabilities for Alice (red) and Bob (green) are shown when each of them plays with their best strategies. The probability of a draw, where neither of them wins, is indicated in gray. The shaded area shows the values of $\theta$ for which the strategies are pure. The experimental results have been plotted on top of the theoretical values.}
  \label{fig:Best_Strategy}
\end{figure}

In the classical game, it is possible to achieve a Nash equilibrium with a mixed strategy where each player wins half the time by choosing 0 or 1 coins at random. However, the deformation $\theta$ allows Alice to have an advantage over Bob when both playing randomly because she can make guesses that increase her winning probability or the draw probability as shown in Figure~\ref{fig:Strategy_random}. In general, these strategies are not stable, and both Alice and Bob can improve them, provided they know each other's strategies. Although Alice's winning probability is higher than Bob's, she can further increase it, as shown in Figure~\ref{fig:Strategy_a}. Bob can try to improve his strategy as well, but in some cases, he is unable to turn the situation around, and Alice's winning probability remains higher as shown in Figure~\ref{fig:Strategy_b}. In these cases, Alice's strategy is stable, although not optimal.

Figure~\ref{fig:strategies} allows us to verify if there are values of $\theta$ that allow an optimal strategy when playing randomly apart from the classical case. This is possible for $\theta=4\pi/3$ since it is equivalent to the classical case under the exchange of states $|1_\theta\rangle$ and $|2_\theta\rangle$. Remarkably, for $\theta=\pi$, it is also possible to achieve equilibrium by playing randomly, although Alice's winning probability is higher than Bob's. In this case, Alice must guess zero coins and Bob one or two. For other values of theta, Alice and Bob's improved strategies are pure. For instance, in the optimal strategy for $\theta=\pi/3$, Alice consistently plays $0$ coins and guesses $0$, while Bob plays $1$ coin and guesses $1$. In this scenario, Alice's winning probability is 0.46, Bob's winning probability is 0.44, and the probability of a draw is 0.1.  This case highlights the counterintuitive nature of the deformed game, as in classical terms, this strategy would allow Bob to win every time.

\subsection{\label{sec:level2}Optimal strategies}
We have calculated the winning probabilities for each player using the experimental data for the overlaps, demonstrating a strong agreement with the theory (see Figs.~\ref{fig:Best_Strategy} and~\ref{fig:strategies}). One may wonder how the optimal strategy of each player changes depending on the $\theta$ deformation. 
To find these strategies, we utilize an exhaustive search approach. Initially, we define Alice's strategy and identify the strategy for Bob to maximize his winning probability. Then, we check whether the best strategy for Alice, considering Bob's strategy, aligns with our initial choice. If this is the case, these strategies establish a Nash equilibrium; otherwise, we iterate the process by changing Alice's strategy.
It is important to note that these strategies rely on information from both players. Therefore, it would not be possible for either of them to discover these strategies beforehand due to the lack of complete information.
The winning probabilities for each player when following the optimal and stable strategy as a function of $\theta$ are shown in Figure~\ref{fig:Best_Strategy}. These strategies are mixed for $\theta\in[4\pi/9,14\pi/9]$, while for the remaining values, they are pure. The transition between mixed and pure strategies occurs at non-trivial $\theta$ values, these are determined by the construction of the operator $X(\theta)$. Multiple applications of $X(\theta)$ will change the phase relation between the states, which can be seen in the exponent of~\ref{d3}, therefore this transition point is expected to be different with more players.

\section{\label{sec:conclusion}Discussion and Conclusion}
In contrast to the classical game, the deformation $\theta$ gives Alice an advantage because she wins more often than Bob, and Bob cannot do anything to change it. Another significant aspect that sets it apart from the classical case is the possibility of both players losing. When playing optimally, the classical game has been shown to be a zero-sum game. This is not true for the quantum game and is more likely to occur at $\theta=\pi$ where there is a $20\%$ chance that nobody wins. Moreover, the Nash equilibrium outside the classical case is no longer Pareto-optimal except at $\theta=0, \ 4\pi/3$ and $2\pi$.

The search for an optimum strategy using shared resources is closely aligned with other competitive games, such the spectrum scarcity problem~\cite{zabaleta2017quantum}. Our analysis has shown that sometimes pure or mixed strategies are available, but may come with different resource overheads. However, we did not consider these resource costs in this paper and leave it as an open problem for future work.

A natural extension of the current implementation is to a three player game, since the four possible outcomes can be encoded on just two photons. In fact, the number of qubits scales as $\left \lceil{\log_2(M+1)}\right \rceil$ (see Appendix~\ref{SM:GQMG}) where $M$ represents the total number of coins, allowing for the efficient construction of a network of players. Moreover, the selection of basis states in this implementation remains arbitrary; we could have opted for three distinct Bell pairs to encode $| \tilde{0}  \rangle$, $| \tilde{1}  \rangle$ and $| \tilde{2}  \rangle$. Using entangled states with minor modifications to the game, could open the possibility of non-local strategies enabled by quantum steering~\cite{schrodinger1935discussion,Wiseman}. This raises the question whether quantum games can be exploited within the network context~\cite{dey2023quantum}, that gain an advantage from multi-partite resources~\cite{proietti2021NQKD,pickston2023Trident}.

In summary, we have outlined a new construction of the two-player quantum Morra game, providing Alice with a winning advantage over Bob that surpasses the classical game. The deformation operator and theoretical states have been realized within a linear optics setup, achieving high fidelity and low standard deviation with respect to the measurement outcomes, which is less  than $ 2\%$.

This formulation of the quantum Morra game has bridged the classical and quantum domains by incorporating the classical game as a particular instance within the extended quantum game. In analysing the strategies, we have identified a new Nash equilibrium that diverges from the classical game. This has enhanced our understanding of strategies in quantum game theory and provided insight into the realization of new quantum games. Surprisingly, it remains unknown whether the equivalent of a Kakutani fixed-point theorem exists in a complex space, such a discovery would guarantee a Nash equilibrium for all quantum games~\cite{khan2018quantum}. Therefore, further theoretical development in quantum game theory is needed to support the implementation of more complex quantum games.

Finally, by extending the principles and techniques applied in this work to other quantum games, we can potentially  evaluate the robustness and efficiency of quantum communication networks and deepen our understanding of quantum game theory.
\begin{acknowledgments}
This work was supported by the UK Engineering and Physical  Sciences Research Council (Grant Nos. EP/T001011/1.) A.S. is supported by the Spanish Ministry of Science and Innovation under the grant SEV-2016-0597-19-4. G.S. acknowledges the support of the Spanish Research Agency (Agencia Estatal de Investigación) through the grants “IFT Centro de Excelencia Severo Ochoa CEX2020-001007-S” and PID2021-127726NB-I00, funded by MCIN/AEI/10.13039/501100011033 and by ERDF, as well as from the CSIC Research Platform on Quantum Technologies PTI-001.
DC is supported by Perimeter Institute for Theoretical Physics. Research at Perimeter Institute is supported in part by the Government of Canada through the Department of Innovation, Science and Economic Development and by the Province of Ontario through the Ministry of Colleges and Universities.
\end{acknowledgments}

\clearpage
\appendix
\section*{Appendices}
The Appendices are organised as follows: Appendix~\ref{SM:Methods} provides additional details on the experimental setup. Appendix~\ref{SM:Qutrits_to_Qubits} provides a detailed description of the qutrit-to-qubit mapping required for the experimental implementation of the game.
The computation of waveplates parameters and experimental overlaps is covered in Appendix~\ref{SM:Parameters}.
Additionally, the unitary transformation is decomposed into one-qubit gates and CNOT gates in Appendix~\ref{SM:Decomposition}. 
Appendix~\ref{SM:GQMG} generalises the deformed game to $N$ players and $M$ coins.
Finally, Appendix~\ref{SM:Coindeposit} introduces the novel concept of Quantum Deposit, drawing inspiration from the quantum Morra game.

\section{\label{SM:Methods}Methods}
\begin{figure}[t]
   \centering
   \includegraphics[width=\linewidth]{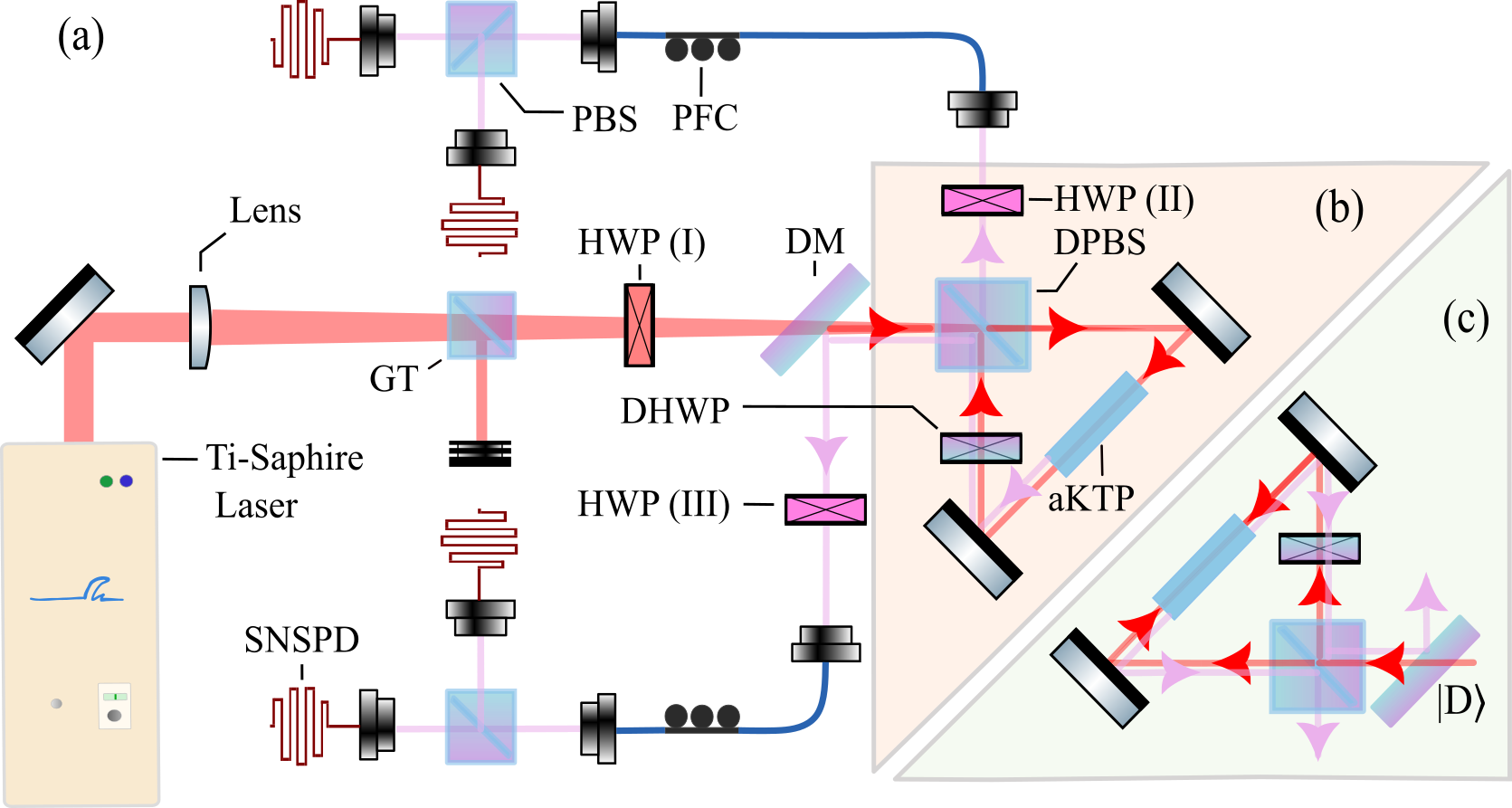}
    \captionsetup{width=1\linewidth}
   \caption{(a) The linear optics circuit, (b) the encoding scheme for $|HH\rangle$ configuration when pumping in Horizontal polarisation and waveplates \{I,II,III\} (c) Pumping with D polarised light. Components: Glan-Taylor (GT), Half Wave Plate (HWP), Dual-wavelength Polarisation Beam Splitter (DPBS), Dichroic Mirror (DM), dual-wavelength HWP (DHWP), Single Mode Fiber (SFM), Polarisation Fiber Controllers (PFC), Polarising Beam Splitter (PBS), Superconducting Nano-wire Single Photon Detectors (SNSPD), aperiodically-polled potassium titanyl phosphate (aKTP).}
   \label{fig:ExpSetup_app}
\end{figure}

We prepare our states using a photon source housed in a Sagnac interferometer to control the PDC process. Figure~\ref{fig:ExpSetup_app} shows a Ti-Sapph laser pumping an aperiodically-polled potassium titanyl phosphate, crystal with 774.8nm light, creating spectrally degenerate photons at 1549.6nm. A lens focuses the beam onto the crystal for optimum PDC events and a Glan-Taylor prism ensures only linearly polarised light enters the Sagnac while a Dichroic Mirror reflects only down converted photons towards the coupler.

The first Half Wave Plate (HWP I) engineered for 775nm and the the Dual-wavelength Polarisation Beam Splitter control the pump direction into transmitted and reflected spatial modes around the Sagnac; Horizontal for clockwise direction and Vertical for Anti-clockwise (AC). A dual-wavelength half-waveplate inside the Sagnc loop rotates V-pol light (In the reflected port) into H-pol to facilitate PDC in the counter-propagating direction. When pumping with D-pol light, the signal and idler photons are indistinguishable and create a maximally entangled state~\eqref{Bell_phase}.
\beq
|\Psi^{+} \rangle = \frac{1}{\sqrt{2}} |h_{s}v_{i}\rangle + e^{i\phi} |v_{s}h_{i}\rangle
\label{Bell_phase}
\eeq
We optimise the phase component from to the fibers and mirrors acting on the two qubits using Polarisation Fiber Controllers (PFC) and recover the $\Psi^{+}$. Down converted signal and idler photons exit the DPBS and are rotated by two 1550nm HWP's (II,III), parameterised by the numerical solution to the Unitary~\eqref{d16} and coupled into a Single Mode Fiber. 

\begin{figure}
    \centering 
    \subfloat[\label{fig:counts1}]{
        \includegraphics[width=0.8\linewidth]{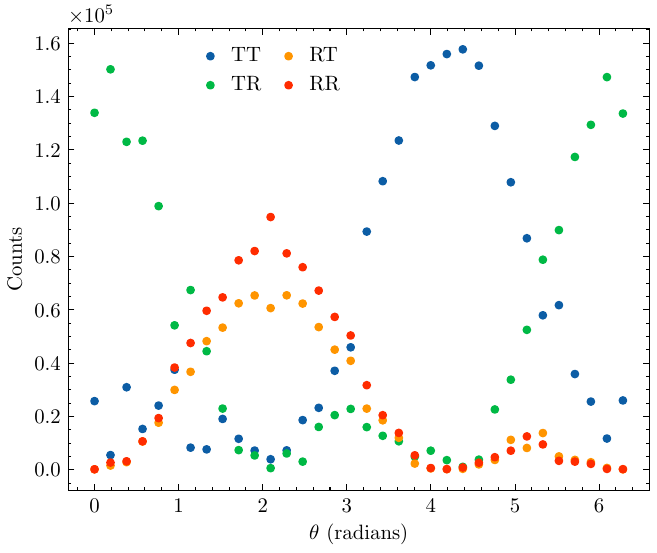} }\\ 
    \vspace{-0.85cm}
    \subfloat[\label{fig:counts2}]{
        \includegraphics[width=0.8\linewidth]{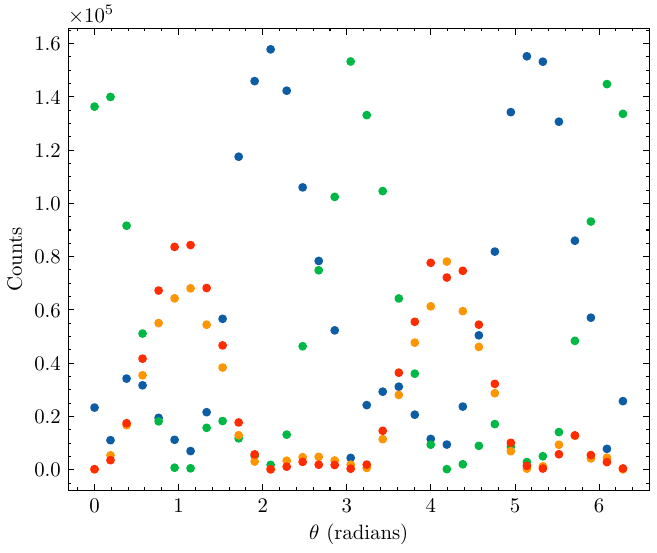} }
    \caption{The coincidence counts on each of the 4 pairs of detectors coupled to the Transmitted (T) and Reflected (R) ports of the PBS to reproduce the overlaps for the states $\left|1_\theta\right\rangle$ \protect\subref{fig:counts1} and $\left|2_\theta\right\rangle$ \protect\subref{fig:counts2}. Error bars have been calculated with Poissonian statistics, but omitted due to size.}
    \label{coinc}
\end{figure}

The single photons detected on SNSPs are counted using a logic box to identify coincidence events in the four detector patters shown in Fig.~\ref{coinc}.
Approximately 150 000 rounds where conducted for 34 evenly spaced values in the interval $[0,2]$.

\section{\label{SM:Qutrits_to_Qubits}From qutrits to qubits}
Let us denote by $|i_A  i_B\rangle, \; (i_A , i_B = 0,1)$ a basis of two qubits states.
The qutrit states will be defined as
\barray 
| \tilde{0}  \rangle & = & |0 0 \rangle \, ,  \label{t1}  \\
| \tilde{1}  \rangle & = &  \frac{1}{\sqrt{2}} ( |0 1 \rangle + | 1 0 \rangle ) \, , \label{t2} \\
| \tilde{2}  \rangle & = & |1 1 \rangle \label{t3}   \, .
\earray 

Equation~\eqref{t9} of the main text expresses $X$ in the qutrit basis $| I \rangle \; (I=  \tilde{0}, \tilde{1}, \tilde{2})$. 
To implement the operation $X$ on two qubit states we have to extend  the basis $| I \rangle$ to a fourth vector
\beq
| \tilde{3} \rangle = \frac{1}{ \sqrt{2}} ( |0 1 \rangle - | 1 0 \rangle )
\label{t10}
\eeq
and promote  $X \rightarrow X_4$  that acts according to
\beq
| \tilde{1}  \rangle = X_4  | \tilde{0} \rangle, \quad 
| \tilde{2}  \rangle = X_4^2  | \tilde{0} \rangle  \, ,
\eeq
and trivially on~\eqref{t10} 
\beq
X_4  | \tilde{3} \rangle =  | \tilde{3} \rangle \, . 
\label{t11}
\eeq
The  matrix  $X_4$ reads in the basis $| I \rangle \; (I= \tilde{0}, \tilde{1}, \tilde{2}, \tilde{3}$)
\beq
X_4 = \left( 
\begin{array}{cccc}
0 & 0 & 1 & 0 \\
1 & 0  & 0  & 0 \\
0 & 1 & 0  & 0\\ 
0 & 0 & 0  & 1\\ 
\end{array} \right)  \, .
\label{t12}
\eeq
To find the action of $X_4$ on the two qubit states we use the unitary transformation $V$ that expresses the basis $| I \rangle$ in terms of the basis $|i_A i_B\rangle$ 
\beq
|I \rangle = \sum_{i_A  i_B}  V_{i_A i_B , I}  | i_A i_B \rangle, \qquad I = \tilde{0}, \tilde{1}, \tilde{2}, \tilde{3} 
\label{t13}
\eeq
where 
\beq
 V = \left( 
 \begin{array}{cccc} 
 1 & 0 & 0 & 0 \\
 0 & \frac{1}{\sqrt{2}} & 0 & \frac{1}{\sqrt{2}}     \\
 0 & \frac{1}{\sqrt{2}} & 0 & -  \frac{1}{\sqrt{2}}    \\
 0 & 0 & 1 & 0 \\
  \end{array} 
 \right)  \, . 
 \label{t14}
 \eeq
The rows are labelled by $00, 01, 10, 11$ and the columns by  $\tilde{0}, \tilde{1}, \tilde{2}, \tilde{3}$. 
The matrix expressing the action of $X_4$ in the two qubit basis is given by 
\beq
 X_{2 \times 2} =  V X_4 V^\dagger =  \left( 
 \begin{array}{cccc} 
 0 & 0 & 0 & 1 \\
 \frac{1}{\sqrt{2}} & \frac{1}{{2}} & - \frac{1}{2} & 0    \\
 \frac{1}{\sqrt{2}} & - \frac{1}{{2}} & \frac{1}{2} & 0    \\
 0 & \frac{1}{\sqrt{2}} & \frac{1}{\sqrt{2}} & 0 \\
  \end{array} 
 \right)  \ . 
 \label{t15}
 \eeq
Similarly, to find the action of the deformed unitary $X_\theta$ on two qubits we first let it act trivially on the state $|\tilde{3} \rangle$, as in~\eqref{t11}, and perform 
the change of basis~\eqref{t13} obtaining $X_{2 \times 2} \rightarrow X_{\theta}$,
\beq
X_{\theta} =
\left( \begin{array}{cccc}
x_0(\theta) & \frac{1}{\sqrt{2}} x_2(\theta) & \frac{x_2(\theta)}{\sqrt{2}}   & x_1(\theta)  \\
\frac{x_1(\theta)}{\sqrt{2}}   & \frac{1}{2} ( 1 + x_0(\theta)) & \frac{1}{2} ( -1 + x_0(\theta) ) &  \frac{x_2(\theta)}{\sqrt{2}}    \\
\frac{x_1(\theta)}{\sqrt{2}}   & \frac{1}{2} ( -1 + x_0(\theta)) & \frac{1}{2} ( 1 + x_0(\theta) ) &  \frac{x_2(\theta)}{\sqrt{2}}   \\
x_2(\theta) & \frac{x_1(\theta)}{\sqrt{2}}  & \frac{x_1(\theta)}{\sqrt{2}}   & x_0(\theta)  
\end{array}
\right)  \,
\label{d16} 
\eeq
This operator gives the states $| 1_\theta \rangle$ and  $| 2_\theta \rangle$ in the two qubit basis,
\barray 
| 1_\theta \rangle & = & x_0(\theta) | 00  \rangle + 
 \frac{x_1(\theta)}{ \sqrt{2}}  ( | 01 \rangle + |10 \rangle )  +  x_2(\theta) | 11  \rangle  \, ,    \label{d9} \\
| 2_\theta \rangle & = & x_0(2 \theta) | 00  \rangle + 
 \frac{x_1(2 \theta)}{ \sqrt{2}}  ( | 01 \rangle + |10 \rangle )  +  x_2(2 \theta) | 11  \rangle  \, .  \nonumber  
 \earray 

\section{\label{SM:Parameters}Parameterised Waveplates}
\begin{figure*}[t]\label{Prob_measure}
\centering
    \subfloat[\label{fig:Prob_overlaps0}]{
        \includegraphics[width=0.33\textwidth]{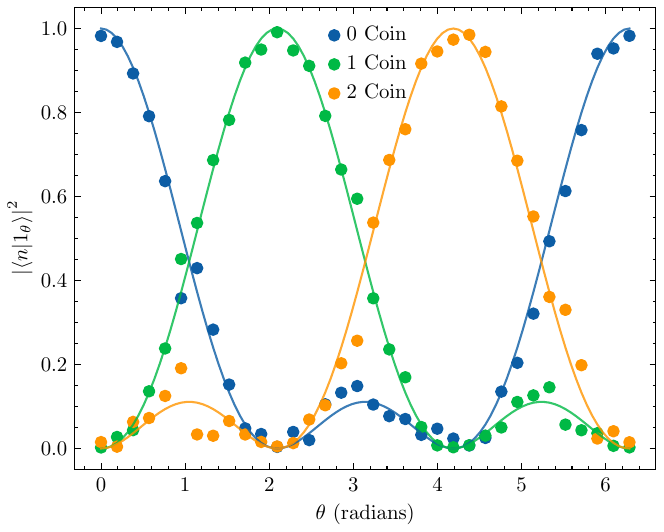}}
    \subfloat[\label{fig:Prob_overlaps1}]{{
        \includegraphics[width=0.33\textwidth]{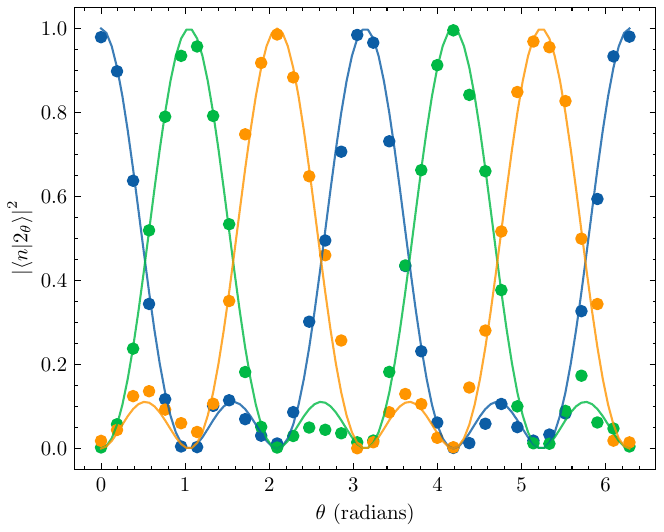}}}
    \subfloat[\label{fig:Entropies}]{{
        \includegraphics[width=0.33\textwidth]{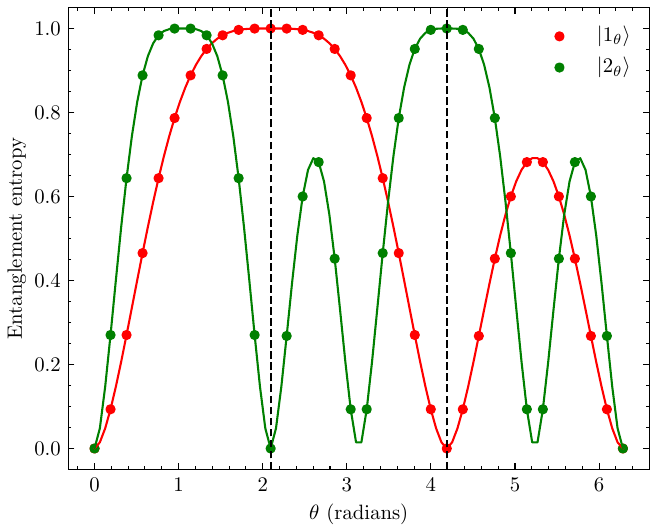}}}
\caption{Features of the states reproduced by our experimental setup. The theoretical (solid line) and experimental (dots) overlaps $\left|\left\langle n|1_\theta\right\rangle\right|^2$ and 
$\left|\left\langle n|2_\theta\right\rangle\right|^2$ with $n=0\text{ (blue)}, 1\text{ (green)}, 2\text{ (orange)}$ are shown in \protect\subref{fig:Prob_overlaps0} and \protect\subref{fig:Prob_overlaps1}, respectively. The entanglement entropy of the states $\left|1_\theta\right\rangle$ and $\left|2_\theta\right\rangle$ is shown in  \protect\subref{fig:Entropies}. The entanglement entropy calculated by simulating the experimental setup without noise and using the optimal parameters for the half-wave plates after numerical optimization is indicated by dots.}
\end{figure*}
The purpose of the experiment is to reproduce the probabilities $p_{a,b}(n)=|\langle n|\psi_{a,b}\rangle |^2$. Therefore, it is not necessary to reproduce the states $\left |\psi_{a,b}\right\rangle$, but only to match the modulus of the coefficients after projecting the state in the $Z$-basis, $q_{i,j}=\left|\left\langle i,j | \psi_{a,b}\right\rangle\right|^2$ with $i=0,1$ and $j=0,1$.

The experimental states $\left |\psi_{a,b}^{exp}\right\rangle$ produced by our setup depend on three parameters, $\vec{\alpha}=(\alpha_1,\alpha_2,\alpha_3)$, through the half-waveplates shown in Figure~\ref{fig:ExpSetup_app}. We need to find the parameters $\vec{\alpha}_{opt}$ that best reproduce $q_{i,j}$. We simulate a noiseless version of the experimental setup obtaining the parameterized states $\left |\psi_{a,b}^{sim}(\vec{\alpha})\right\rangle$ and the overlaps $q_{i,j}^{sim}(\vec{\alpha})$. This allows us to get $\vec{\alpha}_{opt}$ as the solution to an optimization problem in which we aim to maximize the Hellinger fidelity $F$ between $q_{i,j}$ and $q_{i,j}^{sim}(\vec{\alpha})$,
\begin{equation}
    F(\vec{\alpha}) = \sum_{i,j}\sqrt{q_{i,j}q_{i,j}^{sim}(\vec{\alpha})}.
\end{equation}
In order to reproduce the entanglement of the system, we also require that the entanglement entropies of the simulated states $S^{sim}(\vec{\alpha})$ reproduce the theoretical entropies $S$ (see Figure~\ref{fig:Entropies}). Once again, this can be formulated as an optimization problem where the objective is to minimize $\left|S - S^{sim}(\vec{\alpha})\right|$.
Taking into account this new term, we seek to minimize
\begin{equation}
    C(\vec{\alpha}) = 1-F(\vec{\alpha})+\left|S - S^{sim}(\vec{\alpha})\right|,
\end{equation}
so that $\vec{\alpha}_{opt}=\operatorname{argmin}\left(C(\vec{\alpha})\right)$. 
The minimization is performed using the \textit{L-BFGS-B} algorithm~\cite{byrd_limited_1995} implemented in \textsc{SciPy}~\cite{2020SciPy-NMeth}, reaching $C(\vec{\alpha}_{opt}) < 10^{-6}$ for all values of $\theta$.

\section{\label{SM:Decomposition}Unitary decomposition}

An arbitrary unitary $V\in U(4)$ can be implemented as a quantum circuit using local gates and three CNOTs~\cite{vidal_universal_2004}. Furthermore, it is possible to use only two CNOTs iff $\operatorname{tr}\left[V (\sigma_y)^{\otimes 2}V^t(\sigma_y)^{\otimes 2}\right]/\sqrt{\operatorname{det}(V)}$ is real~\cite{shende_recognizing_2004}. This condition is satisfied for the deformed unitary~\eqref{d16} when $\theta=2\pi/3 \ \text{(classical case)}, \ 4\pi/3$. As indicated in the main text, $\theta = 4\pi/3$ reproduces the classical case under the interchange of states $|1\rangle$ and $|2\rangle$. This means that the $\theta$-deformation introduces an additional CNOT gate.

The circuits

\begin{equation}
\scalebox{0.7}{
\begin{quantikz}
&\gate{U(\beta_2,\pi/4,-\pi/3)}&\ctrl{1}&\gate{R_X(-\pi/3)}&\ctrl{1}&\gate{U(\beta_1,-2\pi/3,\pi/4)} \\
&\gate{U(\beta_1,3\pi/4,2\pi/3)}&\targ{}&\gate{R_Z(\pi/3)}&\targ{}&\gate{U(\beta_2,\pi/3,3\pi/4)}
\end{quantikz}
}
\end{equation}

\begin{equation}
\scalebox{0.7}{
\begin{quantikz}
&\gate{U(\beta_2,-\pi/4,-2\pi/3)}&\ctrl{1}&\gate{R_X(-\pi/3)}&\ctrl{1}&\gate{U(\beta_1,-\pi/3,-\pi/4)} \\
&\gate{U(\beta_1,-3\pi/4,\pi/3)}&\targ{}&\gate{R_Z(\pi/3)}&\targ{}&\gate{U(\beta_2,2\pi/3,-3\pi/4)}
\end{quantikz}
}
\end{equation}

with $\beta_1=\arccos(-1/3)/2$ and $\beta_2=\pi-\beta_1$ implement~\eqref{d16} for $\theta = 2\pi/3, \ 4\pi/3$, respectively. For the remaining values of $\theta$, the circuits have the structure 

\begin{equation}
\scalebox{0.7}{
\begin{quantikz}
&\gate{U(\vec{\alpha}_1)}&\ctrl{1}&\gate{U(\vec{\alpha}_3)}&\ctrl{1}&\gate{U(\vec{\alpha}_5)}&\ctrl{1}&\gate{U(\vec{\alpha}_7)} \\
&\gate{U(\vec{\alpha}_2)}&\targ{}&\gate{U(\vec{\alpha}_4)}&\targ{}&\gate{U(\vec{\alpha}_6)}&\targ{}&\gate{U(\vec{\alpha}_8)}
\end{quantikz} 
}
\end{equation}

and the parameters $\vec{\alpha}_j$ can be determined following the method proposed in~\cite{vidal_universal_2004}.

\section{\label{SM:GQMG}General Quantum Morra Game}

The game can be generalized to $n$ players and $m$ coins straightforwardly. Given a pool with $r=nm$ coins, each player can draw from $0$ to $m$ coins and guess the total number of coins, with the restriction that Bob cannot repeat Alice's guess. Similarly to the game with two players and one coin, we assign a state $| \tilde{j}\rangle$ to each possible total number of coins $j=0,1,\ldots,r$. In the classical case, each player has an operator $X_r$ that adds a coin when is applied to a state $\left | \tilde{j}\right\rangle$ with $j<r$, and $X_r^{r+1}=\mathbb{I}$. This operator is the $r+1$-dimensional generalization of the Pauli matrix $\sigma_1$,
\begin{equation}  
X_r = \left (
\begin{array}{cccccc}
0      & 0      & 0      & \cdots & 0     & 1\\
1      & 0      & 0      & \cdots & 0     & 0\\
0      & 1      & 0      & \cdots & 0     & 0\\
0      & 0      & 1      & \cdots & 0     & 0\\
\vdots & \vdots & \vdots & \ddots &\vdots &\vdots\\
0      & 0      & 0      & \cdots & 1     & 0\\ 
\end{array} \right ), 
\end{equation}

and it is related to the operator $Z_r=\operatorname{diag}(1,\omega,\ldots,\omega^{r})$ with $\omega = e^{i2\pi/(r+1)}$ by means of the Fourier transform ${\cal F}$,
\begin{equation}
    X_r = {\cal F}^\dagger Z_r {\cal F} \, .
\end{equation}
Replacing $Z_r$ by $\operatorname{diag}(1,e^{i \theta},\ldots,e^{r i\theta})$ we get the deformed game and the modified $X_{r,\theta}$ operator which creates the states
\begin{equation}
| j_{\theta} \rangle = X^j_{r,\theta} | \tilde{0}  \rangle = x_0(j\theta) | \tilde{0}  \rangle + x_1(j\theta) | \tilde{1}  \rangle + \ldots +  x_r(j\theta) | \tilde{r}  \rangle
\end{equation}
with
\begin{equation}
    x_a(\theta) = \frac{1}{r+1} \left (  1 + \sum_{k=1}^{r}\omega^{(r+1-k) a} e^{k i \theta} \right ) \, .
\end{equation}
The generalized quantum Morra game can be implemented with $\left \lceil{\log_2(r+1)}\right \rceil$ qubits.

\section{\label{SM:Coindeposit}A Quantum Coin Deposit}

It is well known that qubits belonging to a Bell pair are indistinguishable, this is a hallmark of entanglement. An intriguing consequence of our choice of encoding with a Bell pair results in the indistinguishably of the coins played by either player. This may have application in anonymous, secure Quantum Deposit system, wherein the choices of the participants remain concealed. This system operates independently of classical communication channels, offering a sharp departure from traditional games like Morra, which rely on some level of communication. The Quantum Deposit model could lead to new security protocols that require untraceable transactions or enhanced privacy. While the exploration of these applications is beyond the scope of this paper, future researchers could delve into the development of quantum cryptographic methods that capitalize on this work.

\bibliography{references}

\begin{thebibliography}{44}%
\makeatletter
\providecommand \@ifxundefined [1]{%
 \@ifx{#1\undefined}
}%
\providecommand \@ifnum [1]{%
 \ifnum #1\expandafter \@firstoftwo
 \else \expandafter \@secondoftwo
 \fi
}%
\providecommand \@ifx [1]{%
 \ifx #1\expandafter \@firstoftwo
 \else \expandafter \@secondoftwo
 \fi
}%
\providecommand \natexlab [1]{#1}%
\providecommand \enquote  [1]{``#1''}%
\providecommand \bibnamefont  [1]{#1}%
\providecommand \bibfnamefont [1]{#1}%
\providecommand \citenamefont [1]{#1}%
\providecommand \href@noop [0]{\@secondoftwo}%
\providecommand \href [0]{\begingroup \@sanitize@url \@href}%
\providecommand \@href[1]{\@@startlink{#1}\@@href}%
\providecommand \@@href[1]{\endgroup#1\@@endlink}%
\providecommand \@sanitize@url [0]{\catcode `\\12\catcode `\$12\catcode `\&12\catcode `\#12\catcode `\^12\catcode `\_12\catcode `\%12\relax}%
\providecommand \@@startlink[1]{}%
\providecommand \@@endlink[0]{}%
\providecommand \url  [0]{\begingroup\@sanitize@url \@url }%
\providecommand \@url [1]{\endgroup\@href {#1}{\urlprefix }}%
\providecommand \urlprefix  [0]{URL }%
\providecommand \Eprint [0]{\href }%
\providecommand \doibase [0]{http://dx.doi.org/}%
\providecommand \selectlanguage [0]{\@gobble}%
\providecommand \bibinfo  [0]{\@secondoftwo}%
\providecommand \bibfield  [0]{\@secondoftwo}%
\providecommand \translation [1]{[#1]}%
\providecommand \BibitemOpen [0]{}%
\providecommand \bibitemStop [0]{}%
\providecommand \bibitemNoStop [0]{.\EOS\space}%
\providecommand \EOS [0]{\spacefactor3000\relax}%
\providecommand \BibitemShut  [1]{\csname bibitem#1\endcsname}%
\let\auto@bib@innerbib\@empty
\bibitem [{\citenamefont {Von~Neumann}\ and\ \citenamefont {Morgenstern}(1947)}]{von1947theory}%
  \BibitemOpen
  \bibfield  {author} {\bibinfo {author} {\bibfnamefont {John.}\ \bibnamefont {Von~Neumann}}\ and\ \bibinfo {author} {\bibfnamefont {Oskar}\ \bibnamefont {Morgenstern}},\ }\href@noop {} {\emph {\bibinfo {title} {Theory of games and economic behavior, 2nd rev. ed.}}}\ (\bibinfo  {publisher} {Princeton university press},\ \bibinfo {year} {1947})\BibitemShut {NoStop}%
\bibitem [{\citenamefont {Nash}(1950)}]{nash}%
  \BibitemOpen
  \bibfield  {author} {\bibinfo {author} {\bibfnamefont {John~F.}\ \bibnamefont {Nash}},\ }\bibfield  {title} {\enquote {\bibinfo {title} {Equilibrium points in \textit{n} -person games},}\ }\href {\doibase 10.1073/PNAS.36.1.48} {\bibfield  {journal} {\bibinfo  {journal} {Proceedings of the National Academy of Sciences}\ }\textbf {\bibinfo {volume} {36}},\ \bibinfo {pages} {48--49} (\bibinfo {year} {1950})}\BibitemShut {NoStop}%
\bibitem [{\citenamefont {Ghirlanda}\ and\ \citenamefont {Vallortigara}(2004)}]{ghirlanda2004evolution}%
  \BibitemOpen
  \bibfield  {author} {\bibinfo {author} {\bibfnamefont {Stefano}\ \bibnamefont {Ghirlanda}}\ and\ \bibinfo {author} {\bibfnamefont {Giorgio}\ \bibnamefont {Vallortigara}},\ }\bibfield  {title} {\enquote {\bibinfo {title} {The evolution of brain lateralization: a game-theoretical analysis of population structure},}\ }\href@noop {} {\bibfield  {journal} {\bibinfo  {journal} {Proceedings of the Royal Society of London. Series B: Biological Sciences}\ }\textbf {\bibinfo {volume} {271}},\ \bibinfo {pages} {853--857} (\bibinfo {year} {2004})}\BibitemShut {NoStop}%
\bibitem [{\citenamefont {Snidal}(1985)}]{politics}%
  \BibitemOpen
  \bibfield  {author} {\bibinfo {author} {\bibfnamefont {Duncan}\ \bibnamefont {Snidal}},\ }\bibfield  {title} {\enquote {\bibinfo {title} {The {Game} {Theory} of {International} {Politics}},}\ }\href {\doibase 10.2307/2010350} {\bibfield  {journal} {\bibinfo  {journal} {World Politics}\ }\textbf {\bibinfo {volume} {38}},\ \bibinfo {pages} {25--57} (\bibinfo {year} {1985})}\BibitemShut {NoStop}%
\bibitem [{\citenamefont {Luce}\ and\ \citenamefont {Raiffa}(1989)}]{choice}%
  \BibitemOpen
  \bibfield  {author} {\bibinfo {author} {\bibfnamefont {R.~Duncan}\ \bibnamefont {Luce}}\ and\ \bibinfo {author} {\bibfnamefont {Howard}\ \bibnamefont {Raiffa}},\ }\href@noop {} {\emph {\bibinfo {title} {Games and decisions: introduction and critical survey}}}\ (\bibinfo  {publisher} {Dover Publications},\ \bibinfo {address} {New York},\ \bibinfo {year} {1989})\BibitemShut {NoStop}%
\bibitem [{\citenamefont {Shoham}(2008)}]{shoham2008computer}%
  \BibitemOpen
  \bibfield  {author} {\bibinfo {author} {\bibfnamefont {Yoav}\ \bibnamefont {Shoham}},\ }\bibfield  {title} {\enquote {\bibinfo {title} {Computer science and game theory},}\ }\href@noop {} {\bibfield  {journal} {\bibinfo  {journal} {Communications of the ACM}\ }\textbf {\bibinfo {volume} {51}},\ \bibinfo {pages} {74--79} (\bibinfo {year} {2008})}\BibitemShut {NoStop}%
\bibitem [{\citenamefont {Meyer}(1999)}]{Meyer}%
  \BibitemOpen
  \bibfield  {author} {\bibinfo {author} {\bibfnamefont {David~A.}\ \bibnamefont {Meyer}},\ }\bibfield  {title} {\enquote {\bibinfo {title} {Quantum {Strategies}},}\ }\href {\doibase 10.1103/PhysRevLett.82.1052} {\bibfield  {journal} {\bibinfo  {journal} {Physical Review Letters}\ }\textbf {\bibinfo {volume} {82}},\ \bibinfo {pages} {1052--1055} (\bibinfo {year} {1999})}\BibitemShut {NoStop}%
\bibitem [{\citenamefont {Ingarden}(1976)}]{INGARDEN197643}%
  \BibitemOpen
  \bibfield  {author} {\bibinfo {author} {\bibfnamefont {Roman~S.}\ \bibnamefont {Ingarden}},\ }\bibfield  {title} {\enquote {\bibinfo {title} {Quantum information theory},}\ }\href {\doibase https://doi.org/10.1016/0034-4877(76)90005-7} {\bibfield  {journal} {\bibinfo  {journal} {Reports on Mathematical Physics}\ }\textbf {\bibinfo {volume} {10}},\ \bibinfo {pages} {43--72} (\bibinfo {year} {1976})}\BibitemShut {NoStop}%
\bibitem [{\citenamefont {Bennett}\ and\ \citenamefont {Brassard}(2014)}]{Bennett2014}%
  \BibitemOpen
  \bibfield  {author} {\bibinfo {author} {\bibfnamefont {Charles~H.}\ \bibnamefont {Bennett}}\ and\ \bibinfo {author} {\bibfnamefont {Gilles}\ \bibnamefont {Brassard}},\ }\bibfield  {title} {\enquote {\bibinfo {title} {Quantum cryptography: Public key distribution and coin tossing},}\ }\href {\doibase 10.1016/j.tcs.2014.05.025} {\bibfield  {journal} {\bibinfo  {journal} {Theoretical Computer Science}\ }\textbf {\bibinfo {volume} {560}},\ \bibinfo {pages} {7--11} (\bibinfo {year} {2014})}\BibitemShut {NoStop}%
\bibitem [{\citenamefont {Wiesner}(1983)}]{Wiesner}%
  \BibitemOpen
  \bibfield  {author} {\bibinfo {author} {\bibfnamefont {Stephen}\ \bibnamefont {Wiesner}},\ }\bibfield  {title} {\enquote {\bibinfo {title} {Conjugate coding},}\ }\href {\doibase 10.1145/1008908.1008920} {\bibfield  {journal} {\bibinfo  {journal} {ACM SIGACT News}\ }\textbf {\bibinfo {volume} {15}},\ \bibinfo {pages} {78--88} (\bibinfo {year} {1983})}\BibitemShut {NoStop}%
\bibitem [{\citenamefont {Zabaleta}\ \emph {et~al.}(2017)\citenamefont {Zabaleta}, \citenamefont {Barrang{\'u}},\ and\ \citenamefont {Arizmendi}}]{zabaleta2017quantum}%
  \BibitemOpen
  \bibfield  {author} {\bibinfo {author} {\bibfnamefont {Omar~Gustavo}\ \bibnamefont {Zabaleta}}, \bibinfo {author} {\bibfnamefont {Juan~Pablo}\ \bibnamefont {Barrang{\'u}}}, \ and\ \bibinfo {author} {\bibfnamefont {Constancio~M}\ \bibnamefont {Arizmendi}},\ }\bibfield  {title} {\enquote {\bibinfo {title} {Quantum game application to spectrum scarcity problems},}\ }\href {\doibase 10.1016/j.physa.2016.09.054} {\bibfield  {journal} {\bibinfo  {journal} {Physica A: Statistical Mechanics and its Applications}\ }\textbf {\bibinfo {volume} {466}},\ \bibinfo {pages} {455--461} (\bibinfo {year} {2017})}\BibitemShut {NoStop}%
\bibitem [{\citenamefont {Eisert}\ \emph {et~al.}(1999)\citenamefont {Eisert}, \citenamefont {Wilkens},\ and\ \citenamefont {Lewenstein}}]{Eisert}%
  \BibitemOpen
  \bibfield  {author} {\bibinfo {author} {\bibfnamefont {Jens}\ \bibnamefont {Eisert}}, \bibinfo {author} {\bibfnamefont {Martin}\ \bibnamefont {Wilkens}}, \ and\ \bibinfo {author} {\bibfnamefont {Maciej}\ \bibnamefont {Lewenstein}},\ }\bibfield  {title} {\enquote {\bibinfo {title} {Quantum {Games} and {Quantum} {Strategies}},}\ }\href {\doibase 10.1103/PhysRevLett.83.3077} {\bibfield  {journal} {\bibinfo  {journal} {Physical Review Letters}\ }\textbf {\bibinfo {volume} {83}},\ \bibinfo {pages} {3077--3080} (\bibinfo {year} {1999})}\BibitemShut {NoStop}%
\bibitem [{\citenamefont {Benjamin}\ and\ \citenamefont {Hayden}(2001)}]{benjamin2001multiplayer}%
  \BibitemOpen
  \bibfield  {author} {\bibinfo {author} {\bibfnamefont {Simon~C}\ \bibnamefont {Benjamin}}\ and\ \bibinfo {author} {\bibfnamefont {Patrick~M}\ \bibnamefont {Hayden}},\ }\bibfield  {title} {\enquote {\bibinfo {title} {Multiplayer quantum games},}\ }\href {\doibase 10.1103/PhysRevA.64.030301} {\bibfield  {journal} {\bibinfo  {journal} {Physical Review A}\ }\textbf {\bibinfo {volume} {64}},\ \bibinfo {pages} {030301(R)} (\bibinfo {year} {2001})}\BibitemShut {NoStop}%
\bibitem [{\citenamefont {Anand}\ \emph {et~al.}(2020)\citenamefont {Anand}, \citenamefont {Behera},\ and\ \citenamefont {Panigrahi}}]{anand2020prisonerIBM}%
  \BibitemOpen
  \bibfield  {author} {\bibinfo {author} {\bibfnamefont {Amit}\ \bibnamefont {Anand}}, \bibinfo {author} {\bibfnamefont {Bikash~K}\ \bibnamefont {Behera}}, \ and\ \bibinfo {author} {\bibfnamefont {Prasanta~K}\ \bibnamefont {Panigrahi}},\ }\bibfield  {title} {\enquote {\bibinfo {title} {Solving diner’s dilemma game, circuit implementation and verification on the ibm quantum simulator},}\ }\href {\doibase 10.1007/s11128-020-02687-5} {\bibfield  {journal} {\bibinfo  {journal} {Quantum Information Processing}\ }\textbf {\bibinfo {volume} {19}},\ \bibinfo {pages} {1--14} (\bibinfo {year} {2020})}\BibitemShut {NoStop}%
\bibitem [{\citenamefont {Solmeyer}\ \emph {et~al.}(2018)\citenamefont {Solmeyer}, \citenamefont {Linke}, \citenamefont {Figgatt}, \citenamefont {Landsman}, \citenamefont {Balu}, \citenamefont {Siopsis},\ and\ \citenamefont {Monroe}}]{solmeyer2018demonstration}%
  \BibitemOpen
  \bibfield  {author} {\bibinfo {author} {\bibfnamefont {Neal}\ \bibnamefont {Solmeyer}}, \bibinfo {author} {\bibfnamefont {Norbert~M}\ \bibnamefont {Linke}}, \bibinfo {author} {\bibfnamefont {Caroline}\ \bibnamefont {Figgatt}}, \bibinfo {author} {\bibfnamefont {Kevin~A}\ \bibnamefont {Landsman}}, \bibinfo {author} {\bibfnamefont {Radhakrishnan}\ \bibnamefont {Balu}}, \bibinfo {author} {\bibfnamefont {George}\ \bibnamefont {Siopsis}}, \ and\ \bibinfo {author} {\bibfnamefont {Christopher}\ \bibnamefont {Monroe}},\ }\bibfield  {title} {\enquote {\bibinfo {title} {Demonstration of a bayesian quantum game on an ion-trap quantum computer},}\ }\href {\doibase 10.1088/2058-9565/aacf0e} {\bibfield  {journal} {\bibinfo  {journal} {Quantum Science and Technology}\ }\textbf {\bibinfo {volume} {3}},\ \bibinfo {pages} {045002} (\bibinfo {year} {2018})}\BibitemShut {NoStop}%
\bibitem [{\citenamefont {George}\ \emph {et~al.}(2013)\citenamefont {George}, \citenamefont {Robledo}, \citenamefont {Maroney}, \citenamefont {Blok}, \citenamefont {Bernien}, \citenamefont {Markham}, \citenamefont {Twitchen}, \citenamefont {Morton}, \citenamefont {Briggs},\ and\ \citenamefont {Hanson}}]{george2013threeboxes}%
  \BibitemOpen
  \bibfield  {author} {\bibinfo {author} {\bibfnamefont {Richard~E}\ \bibnamefont {George}}, \bibinfo {author} {\bibfnamefont {Lucio~M}\ \bibnamefont {Robledo}}, \bibinfo {author} {\bibfnamefont {Owen~JE}\ \bibnamefont {Maroney}}, \bibinfo {author} {\bibfnamefont {Machiel~S}\ \bibnamefont {Blok}}, \bibinfo {author} {\bibfnamefont {Hannes}\ \bibnamefont {Bernien}}, \bibinfo {author} {\bibfnamefont {Matthew~L}\ \bibnamefont {Markham}}, \bibinfo {author} {\bibfnamefont {Daniel~J}\ \bibnamefont {Twitchen}}, \bibinfo {author} {\bibfnamefont {John~JL}\ \bibnamefont {Morton}}, \bibinfo {author} {\bibfnamefont {G~Andrew~D}\ \bibnamefont {Briggs}}, \ and\ \bibinfo {author} {\bibfnamefont {Ronald}\ \bibnamefont {Hanson}},\ }\bibfield  {title} {\enquote {\bibinfo {title} {Opening up three quantum boxes causes classically undetectable wavefunction collapse},}\ }\href {\doibase 10.1073/pnas.1208374110} {\bibfield  {journal} {\bibinfo  {journal} {Proceedings of the National Academy of Sciences}\ }\textbf {\bibinfo {volume}
  {110}},\ \bibinfo {pages} {3777--3781} (\bibinfo {year} {2013})}\BibitemShut {NoStop}%
\bibitem [{\citenamefont {Prevedel}\ \emph {et~al.}(2007)\citenamefont {Prevedel}, \citenamefont {Stefanov}, \citenamefont {Walther},\ and\ \citenamefont {Zeilinger}}]{zeilinger2007prisoner}%
  \BibitemOpen
  \bibfield  {author} {\bibinfo {author} {\bibfnamefont {Robert}\ \bibnamefont {Prevedel}}, \bibinfo {author} {\bibfnamefont {Andr{\'e}}\ \bibnamefont {Stefanov}}, \bibinfo {author} {\bibfnamefont {Philip}\ \bibnamefont {Walther}}, \ and\ \bibinfo {author} {\bibfnamefont {Anton}\ \bibnamefont {Zeilinger}},\ }\bibfield  {title} {\enquote {\bibinfo {title} {Experimental realization of a quantum game on a one-way quantum computer},}\ }\href {\doibase 10.1088/1367-2630/9/6/205} {\bibfield  {journal} {\bibinfo  {journal} {New Journal of Physics}\ }\textbf {\bibinfo {volume} {9}},\ \bibinfo {pages} {205} (\bibinfo {year} {2007})}\BibitemShut {NoStop}%
\bibitem [{\citenamefont {Pinheiro}\ \emph {et~al.}(2013)\citenamefont {Pinheiro}, \citenamefont {Souza}, \citenamefont {Caetano}, \citenamefont {Huguenin}, \citenamefont {Schmidt},\ and\ \citenamefont {Khoury}}]{pinheiro2013vectorvortex}%
  \BibitemOpen
  \bibfield  {author} {\bibinfo {author} {\bibfnamefont {ARC}\ \bibnamefont {Pinheiro}}, \bibinfo {author} {\bibfnamefont {CER}\ \bibnamefont {Souza}}, \bibinfo {author} {\bibfnamefont {DP}~\bibnamefont {Caetano}}, \bibinfo {author} {\bibfnamefont {JAO}\ \bibnamefont {Huguenin}}, \bibinfo {author} {\bibfnamefont {AGM}\ \bibnamefont {Schmidt}}, \ and\ \bibinfo {author} {\bibfnamefont {AZ}~\bibnamefont {Khoury}},\ }\bibfield  {title} {\enquote {\bibinfo {title} {Vector vortex implementation of a quantum game},}\ }\href {\doibase 10.1364/JOSAB.30.003210} {\bibfield  {journal} {\bibinfo  {journal} {JOSA B}\ }\textbf {\bibinfo {volume} {30}},\ \bibinfo {pages} {3210--3214} (\bibinfo {year} {2013})}\BibitemShut {NoStop}%
\bibitem [{\citenamefont {Schmid}\ \emph {et~al.}(2010)\citenamefont {Schmid}, \citenamefont {Flitney}, \citenamefont {Wieczorek}, \citenamefont {Kiesel}, \citenamefont {Weinfurter},\ and\ \citenamefont {Hollenberg}}]{schmid2010experimentalminority}%
  \BibitemOpen
  \bibfield  {author} {\bibinfo {author} {\bibfnamefont {Christian}\ \bibnamefont {Schmid}}, \bibinfo {author} {\bibfnamefont {Adrian~P}\ \bibnamefont {Flitney}}, \bibinfo {author} {\bibfnamefont {Witlef}\ \bibnamefont {Wieczorek}}, \bibinfo {author} {\bibfnamefont {Nikolai}\ \bibnamefont {Kiesel}}, \bibinfo {author} {\bibfnamefont {Harald}\ \bibnamefont {Weinfurter}}, \ and\ \bibinfo {author} {\bibfnamefont {Lloyd~CL}\ \bibnamefont {Hollenberg}},\ }\bibfield  {title} {\enquote {\bibinfo {title} {Experimental implementation of a four-player quantum game},}\ }\href {\doibase 10.1088/1367-2630/12/6/063031} {\bibfield  {journal} {\bibinfo  {journal} {New Journal of Physics}\ }\textbf {\bibinfo {volume} {12}},\ \bibinfo {pages} {063031} (\bibinfo {year} {2010})}\BibitemShut {NoStop}%
\bibitem [{\citenamefont {Balthazar}\ \emph {et~al.}(2015)\citenamefont {Balthazar}, \citenamefont {Passos}, \citenamefont {Schmidt}, \citenamefont {Caetano},\ and\ \citenamefont {Huguenin}}]{balthazar_experimental_2015}%
  \BibitemOpen
  \bibfield  {author} {\bibinfo {author} {\bibfnamefont {W~F}\ \bibnamefont {Balthazar}}, \bibinfo {author} {\bibfnamefont {M~H~M}\ \bibnamefont {Passos}}, \bibinfo {author} {\bibfnamefont {A~G~M}\ \bibnamefont {Schmidt}}, \bibinfo {author} {\bibfnamefont {D~P}\ \bibnamefont {Caetano}}, \ and\ \bibinfo {author} {\bibfnamefont {J~A~O}\ \bibnamefont {Huguenin}},\ }\bibfield  {title} {\enquote {\bibinfo {title} {Experimental realization of the quantum duel game using linear optical circuits},}\ }\href {\doibase 10.1088/0953-4075/48/16/165505} {\bibfield  {journal} {\bibinfo  {journal} {Journal of Physics B: Atomic, Molecular and Optical Physics}\ }\textbf {\bibinfo {volume} {48}},\ \bibinfo {pages} {165505} (\bibinfo {year} {2015})}\BibitemShut {NoStop}%
\bibitem [{\citenamefont {Dey}\ \emph {et~al.}(2023)\citenamefont {Dey}, \citenamefont {Marchetti}, \citenamefont {Caleffi},\ and\ \citenamefont {Cacciapuoti}}]{dey2023quantum}%
  \BibitemOpen
  \bibfield  {author} {\bibinfo {author} {\bibfnamefont {Indrakshi}\ \bibnamefont {Dey}}, \bibinfo {author} {\bibfnamefont {Nicola}\ \bibnamefont {Marchetti}}, \bibinfo {author} {\bibfnamefont {Marcello}\ \bibnamefont {Caleffi}}, \ and\ \bibinfo {author} {\bibfnamefont {Angela~Sara}\ \bibnamefont {Cacciapuoti}},\ }\bibfield  {title} {\enquote {\bibinfo {title} {Quantum game theory meets quantum networks},}\ }\href@noop {} {\bibfield  {journal} {\bibinfo  {journal} {arXiv preprint arXiv:2306.08928}\ } (\bibinfo {year} {2023})}\BibitemShut {NoStop}%
\bibitem [{\citenamefont {Suetonius}(2019)}]{suetonius2019lives}%
  \BibitemOpen
  \bibfield  {author} {\bibinfo {author} {\bibfnamefont {Gaius}\ \bibnamefont {Suetonius}},\ }\href@noop {} {\emph {\bibinfo {title} {The lives of the twelve Caesars}}}\ (\bibinfo  {publisher} {BoD--Books on Demand},\ \bibinfo {year} {2019})\BibitemShut {NoStop}%
\bibitem [{\citenamefont {Morris}(2012)}]{morris2012introduction}%
  \BibitemOpen
  \bibfield  {author} {\bibinfo {author} {\bibfnamefont {Peter}\ \bibnamefont {Morris}},\ }\href@noop {} {\emph {\bibinfo {title} {Introduction to game theory}}}\ (\bibinfo  {publisher} {Springer Science \& Business Media},\ \bibinfo {year} {2012})\BibitemShut {NoStop}%
\bibitem [{\citenamefont {Pastor-Abia}\ \emph {et~al.}(2001)\citenamefont {Pastor-Abia}, \citenamefont {Pérez-Jordá}, \citenamefont {San-Fabián}, \citenamefont {Louis},\ and\ \citenamefont {Vega-Redondo}}]{Pastor}%
  \BibitemOpen
  \bibfield  {author} {\bibinfo {author} {\bibfnamefont {Luis}\ \bibnamefont {Pastor-Abia}}, \bibinfo {author} {\bibfnamefont {José~M.}\ \bibnamefont {Pérez-Jordá}}, \bibinfo {author} {\bibfnamefont {Emilio}\ \bibnamefont {San-Fabián}}, \bibinfo {author} {\bibfnamefont {Enrique}\ \bibnamefont {Louis}}, \ and\ \bibinfo {author} {\bibfnamefont {Fernando}\ \bibnamefont {Vega-Redondo}},\ }\bibfield  {title} {\enquote {\bibinfo {title} {Strategic behavior and information transmission in a stylized (so-called \emph{Chinos}) guessing game},}\ }\href {\doibase 10.1142/S0219525901000152} {\bibfield  {journal} {\bibinfo  {journal} {Advances in Complex Systems}\ }\textbf {\bibinfo {volume} {04}},\ \bibinfo {pages} {177--190} (\bibinfo {year} {2001})}\BibitemShut {NoStop}%
\bibitem [{\citenamefont {Pastor-Abia}\ \emph {et~al.}(2003)\citenamefont {Pastor-Abia}, \citenamefont {San-Fabi{\'a}n}, \citenamefont {Louis},\ and\ \citenamefont {Vega-Redondo}}]{pastor2003learning}%
  \BibitemOpen
  \bibfield  {author} {\bibinfo {author} {\bibfnamefont {Luis}\ \bibnamefont {Pastor-Abia}}, \bibinfo {author} {\bibfnamefont {Emilio}\ \bibnamefont {San-Fabi{\'a}n}}, \bibinfo {author} {\bibfnamefont {Enrique}\ \bibnamefont {Louis}}, \ and\ \bibinfo {author} {\bibfnamefont {Fernando}\ \bibnamefont {Vega-Redondo}},\ }\bibfield  {title} {\enquote {\bibinfo {title} {Learning to play in a stylized ({Chinos}) game: some preliminary results},}\ }in\ \href@noop {} {\emph {\bibinfo {booktitle} {AIP Conference Proceedings}}},\ Vol.\ \bibinfo {volume} {661}\ (\bibinfo {organization} {American Institute of Physics},\ \bibinfo {year} {2003})\ pp.\ \bibinfo {pages} {167--173}\BibitemShut {NoStop}%
\bibitem [{\citenamefont {Guinea}\ and\ \citenamefont {Mart{\'i}n-Delgado}(2003)}]{GMD}%
  \BibitemOpen
  \bibfield  {author} {\bibinfo {author} {\bibfnamefont {F}~\bibnamefont {Guinea}}\ and\ \bibinfo {author} {\bibfnamefont {M~A}\ \bibnamefont {Mart{\'i}n-Delgado}},\ }\bibfield  {title} {\enquote {\bibinfo {title} {Quantum {Chinos} game: winning strategies through quantum fluctuations},}\ }\href {\doibase 10.1088/0305-4470/36/13/104} {\bibfield  {journal} {\bibinfo  {journal} {Journal of Physics A: Mathematical and General}\ }\textbf {\bibinfo {volume} {36}},\ \bibinfo {pages} {L197--L204} (\bibinfo {year} {2003})}\BibitemShut {NoStop}%
\bibitem [{\citenamefont {Centeno}\ and\ \citenamefont {Sierra}(2022)}]{centeno2022Chinos}%
  \BibitemOpen
  \bibfield  {author} {\bibinfo {author} {\bibfnamefont {Daniel}\ \bibnamefont {Centeno}}\ and\ \bibinfo {author} {\bibfnamefont {Germ{\'a}n}\ \bibnamefont {Sierra}},\ }\bibfield  {title} {\enquote {\bibinfo {title} {General quantum chinos games},}\ }\href {\doibase 10.1088/2399-6528/ac7434} {\bibfield  {journal} {\bibinfo  {journal} {Journal of Physics Communications}\ }\textbf {\bibinfo {volume} {6}},\ \bibinfo {pages} {075009} (\bibinfo {year} {2022})}\BibitemShut {NoStop}%
\bibitem [{\citenamefont {Glicksberg}(1952)}]{glicksberg1952further}%
  \BibitemOpen
  \bibfield  {author} {\bibinfo {author} {\bibfnamefont {Irving~L}\ \bibnamefont {Glicksberg}},\ }\bibfield  {title} {\enquote {\bibinfo {title} {A further generalization of the kakutani fixed theorem, with application to nash equilibrium points},}\ }\href {\doibase 10.2307/2032478} {\bibfield  {journal} {\bibinfo  {journal} {Proceedings of the American Mathematical Society}\ }\textbf {\bibinfo {volume} {3}},\ \bibinfo {pages} {170--174} (\bibinfo {year} {1952})}\BibitemShut {NoStop}%
\bibitem [{\citenamefont {Bicchieri}(2004)}]{mele_oxford_2004}%
  \BibitemOpen
  \bibfield  {author} {\bibinfo {author} {\bibfnamefont {Cristina}\ \bibnamefont {Bicchieri}},\ }\bibfield  {title} {\enquote {\bibinfo {title} {Rationality and game theory},}\ }in\ \href {\doibase 10.1093/oxfordhb/9780195145397.001.0001} {\emph {\bibinfo {booktitle} {The {Oxford} {Handbook} of {Rationality}}}},\ \bibinfo {editor} {edited by\ \bibinfo {editor} {\bibfnamefont {Alfred~R.}\ \bibnamefont {Mele}}\ and\ \bibinfo {editor} {\bibfnamefont {Piers}\ \bibnamefont {Rawling}}}\ (\bibinfo  {publisher} {Oxford University Press},\ \bibinfo {year} {2004})\ \bibinfo {edition} {1st}\ ed.,\ Chap.~\bibinfo {chapter} {10}, pp.\ \bibinfo {pages} {182--205}\BibitemShut {NoStop}%
\bibitem [{\citenamefont {Askari}\ \emph {et~al.}(2019)\citenamefont {Askari}, \citenamefont {Gordji},\ and\ \citenamefont {Park}}]{askari2019behavioral}%
  \BibitemOpen
  \bibfield  {author} {\bibinfo {author} {\bibfnamefont {Gholamreza}\ \bibnamefont {Askari}}, \bibinfo {author} {\bibfnamefont {Madjid~Eshaghi}\ \bibnamefont {Gordji}}, \ and\ \bibinfo {author} {\bibfnamefont {Choonkil}\ \bibnamefont {Park}},\ }\bibfield  {title} {\enquote {\bibinfo {title} {The behavioral model and game theory},}\ }\href@noop {} {\bibfield  {journal} {\bibinfo  {journal} {Palgrave Communications}\ }\textbf {\bibinfo {volume} {5}} (\bibinfo {year} {2019})}\BibitemShut {NoStop}%
\bibitem [{\citenamefont {Naseer}\ \emph {et~al.}(2021)\citenamefont {Naseer}, \citenamefont {Minhas}, \citenamefont {Saleem}, \citenamefont {Siddiqui}, \citenamefont {Bhatti},\ and\ \citenamefont {Mahmood}}]{naseer2021game}%
  \BibitemOpen
  \bibfield  {author} {\bibinfo {author} {\bibfnamefont {Sundus}\ \bibnamefont {Naseer}}, \bibinfo {author} {\bibfnamefont {Qurratul-Ain}\ \bibnamefont {Minhas}}, \bibinfo {author} {\bibfnamefont {Khalid}\ \bibnamefont {Saleem}}, \bibinfo {author} {\bibfnamefont {Ghazanfar~Farooq}\ \bibnamefont {Siddiqui}}, \bibinfo {author} {\bibfnamefont {Naeem}\ \bibnamefont {Bhatti}}, \ and\ \bibinfo {author} {\bibfnamefont {Hasan}\ \bibnamefont {Mahmood}},\ }\bibfield  {title} {\enquote {\bibinfo {title} {A game theoretic power control and spectrum sharing approach using cost dominance in cognitive radio networks},}\ }\href@noop {} {\bibfield  {journal} {\bibinfo  {journal} {PeerJ Computer Science}\ }\textbf {\bibinfo {volume} {7}},\ \bibinfo {pages} {e617} (\bibinfo {year} {2021})}\BibitemShut {NoStop}%
\bibitem [{\citenamefont {Pareto}(2008)}]{pareto}%
  \BibitemOpen
  \bibfield  {author} {\bibinfo {author} {\bibfnamefont {Vilfredo}\ \bibnamefont {Pareto}},\ }\bibfield  {title} {\enquote {\bibinfo {title} {The maximum of utility given by free competition},}\ }\href@noop {} {\bibfield  {journal} {\bibinfo  {journal} {Giornale degli Economisti e Annali di Economia}\ }\textbf {\bibinfo {volume} {67 (Anno 121)}},\ \bibinfo {pages} {387--403} (\bibinfo {year} {2008})}\BibitemShut {NoStop}%
\bibitem [{\citenamefont {Vidal}\ and\ \citenamefont {Dawson}(2004)}]{vidal_universal_2004}%
  \BibitemOpen
  \bibfield  {author} {\bibinfo {author} {\bibfnamefont {G.}~\bibnamefont {Vidal}}\ and\ \bibinfo {author} {\bibfnamefont {C.~M.}\ \bibnamefont {Dawson}},\ }\bibfield  {title} {\enquote {\bibinfo {title} {Universal quantum circuit for two-qubit transformations with three controlled-{NOT} gates},}\ }\href {\doibase 10.1103/PhysRevA.69.010301} {\bibfield  {journal} {\bibinfo  {journal} {Physical Review A}\ }\textbf {\bibinfo {volume} {69}},\ \bibinfo {pages} {010301(R)} (\bibinfo {year} {2004})}\BibitemShut {NoStop}%
\bibitem [{\citenamefont {Shende}\ \emph {et~al.}(2004)\citenamefont {Shende}, \citenamefont {Bullock},\ and\ \citenamefont {Markov}}]{shende_recognizing_2004}%
  \BibitemOpen
  \bibfield  {author} {\bibinfo {author} {\bibfnamefont {Vivek~V.}\ \bibnamefont {Shende}}, \bibinfo {author} {\bibfnamefont {Stephen~S.}\ \bibnamefont {Bullock}}, \ and\ \bibinfo {author} {\bibfnamefont {Igor~L.}\ \bibnamefont {Markov}},\ }\bibfield  {title} {\enquote {\bibinfo {title} {Recognizing small-circuit structure in two-qubit operators},}\ }\href {\doibase 10.1103/PhysRevA.70.012310} {\bibfield  {journal} {\bibinfo  {journal} {Physical Review A}\ }\textbf {\bibinfo {volume} {70}},\ \bibinfo {pages} {012310} (\bibinfo {year} {2004})}\BibitemShut {NoStop}%
\bibitem [{\citenamefont {Fedrizzi}\ \emph {et~al.}(2007)\citenamefont {Fedrizzi}, \citenamefont {Herbst}, \citenamefont {Poppe}, \citenamefont {Jennewein},\ and\ \citenamefont {Zeilinger}}]{fedrizzi2007wavelength}%
  \BibitemOpen
  \bibfield  {author} {\bibinfo {author} {\bibfnamefont {Alessandro}\ \bibnamefont {Fedrizzi}}, \bibinfo {author} {\bibfnamefont {Thomas}\ \bibnamefont {Herbst}}, \bibinfo {author} {\bibfnamefont {Andreas}\ \bibnamefont {Poppe}}, \bibinfo {author} {\bibfnamefont {Thomas}\ \bibnamefont {Jennewein}}, \ and\ \bibinfo {author} {\bibfnamefont {Anton}\ \bibnamefont {Zeilinger}},\ }\bibfield  {title} {\enquote {\bibinfo {title} {A wavelength-tunable fiber-coupled source of narrowband entangled photons},}\ }\href@noop {} {\bibfield  {journal} {\bibinfo  {journal} {Optics Express}\ }\textbf {\bibinfo {volume} {15}},\ \bibinfo {pages} {15377--15386} (\bibinfo {year} {2007})}\BibitemShut {NoStop}%
\bibitem [{\citenamefont {Pickston}\ \emph {et~al.}(2023)\citenamefont {Pickston}, \citenamefont {Ho}, \citenamefont {Ulibarrena}, \citenamefont {Grasselli}, \citenamefont {Proietti}, \citenamefont {Morrison}, \citenamefont {Barrow}, \citenamefont {Graffitti},\ and\ \citenamefont {Fedrizzi}}]{pickston2023Trident}%
  \BibitemOpen
  \bibfield  {author} {\bibinfo {author} {\bibfnamefont {Alexander}\ \bibnamefont {Pickston}}, \bibinfo {author} {\bibfnamefont {Joseph}\ \bibnamefont {Ho}}, \bibinfo {author} {\bibfnamefont {Andr{\'e}s}\ \bibnamefont {Ulibarrena}}, \bibinfo {author} {\bibfnamefont {Federico}\ \bibnamefont {Grasselli}}, \bibinfo {author} {\bibfnamefont {Massimiliano}\ \bibnamefont {Proietti}}, \bibinfo {author} {\bibfnamefont {Christopher~L}\ \bibnamefont {Morrison}}, \bibinfo {author} {\bibfnamefont {Peter}\ \bibnamefont {Barrow}}, \bibinfo {author} {\bibfnamefont {Francesco}\ \bibnamefont {Graffitti}}, \ and\ \bibinfo {author} {\bibfnamefont {Alessandro}\ \bibnamefont {Fedrizzi}},\ }\bibfield  {title} {\enquote {\bibinfo {title} {Conference key agreement in a quantum network},}\ }\href {\doibase 10.1038/s41534-023-00750-4} {\bibfield  {journal} {\bibinfo  {journal} {npj Quantum Information}\ }\textbf {\bibinfo {volume} {9}},\ \bibinfo {pages} {82} (\bibinfo {year} {2023})}\BibitemShut {NoStop}%
\bibitem [{\citenamefont {Byrd}\ \emph {et~al.}(1995)\citenamefont {Byrd}, \citenamefont {Lu}, \citenamefont {Nocedal},\ and\ \citenamefont {Zhu}}]{byrd_limited_1995}%
  \BibitemOpen
  \bibfield  {author} {\bibinfo {author} {\bibfnamefont {Richard~H.}\ \bibnamefont {Byrd}}, \bibinfo {author} {\bibfnamefont {Peihuang}\ \bibnamefont {Lu}}, \bibinfo {author} {\bibfnamefont {Jorge}\ \bibnamefont {Nocedal}}, \ and\ \bibinfo {author} {\bibfnamefont {Ciyou}\ \bibnamefont {Zhu}},\ }\bibfield  {title} {\enquote {\bibinfo {title} {A {Limited} {Memory} {Algorithm} for {Bound} {Constrained} {Optimization}},}\ }\href {\doibase 10.1137/0916069} {\bibfield  {journal} {\bibinfo  {journal} {SIAM Journal on Scientific Computing}\ }\textbf {\bibinfo {volume} {16}},\ \bibinfo {pages} {1190--1208} (\bibinfo {year} {1995})}\BibitemShut {NoStop}%
\bibitem [{\citenamefont {Virtanen}\ \emph {et~al.}(2020)\citenamefont {Virtanen}, \citenamefont {Gommers}, \citenamefont {Oliphant}, \citenamefont {Haberland}, \citenamefont {Reddy}, \citenamefont {Cournapeau}, \citenamefont {Burovski}, \citenamefont {Peterson}, \citenamefont {Weckesser}, \citenamefont {Bright}, \citenamefont {{van der Walt}}, \citenamefont {Brett}, \citenamefont {Wilson}, \citenamefont {Millman}, \citenamefont {Mayorov}, \citenamefont {Nelson}, \citenamefont {Jones}, \citenamefont {Kern}, \citenamefont {Larson}, \citenamefont {Carey}, \citenamefont {Polat}, \citenamefont {Feng}, \citenamefont {Moore}, \citenamefont {{VanderPlas}}, \citenamefont {Laxalde}, \citenamefont {Perktold}, \citenamefont {Cimrman}, \citenamefont {Henriksen}, \citenamefont {Quintero}, \citenamefont {Harris}, \citenamefont {Archibald}, \citenamefont {Ribeiro}, \citenamefont {Pedregosa}, \citenamefont {{van Mulbregt}},\ and\ \citenamefont {{SciPy 1.0 Contributors}}}]{2020SciPy-NMeth}%
  \BibitemOpen
  \bibfield  {author} {\bibinfo {author} {\bibfnamefont {Pauli}\ \bibnamefont {Virtanen}}, \bibinfo {author} {\bibfnamefont {Ralf}\ \bibnamefont {Gommers}}, \bibinfo {author} {\bibfnamefont {Travis~E.}\ \bibnamefont {Oliphant}}, \bibinfo {author} {\bibfnamefont {Matt}\ \bibnamefont {Haberland}}, \bibinfo {author} {\bibfnamefont {Tyler}\ \bibnamefont {Reddy}}, \bibinfo {author} {\bibfnamefont {David}\ \bibnamefont {Cournapeau}}, \bibinfo {author} {\bibfnamefont {Evgeni}\ \bibnamefont {Burovski}}, \bibinfo {author} {\bibfnamefont {Pearu}\ \bibnamefont {Peterson}}, \bibinfo {author} {\bibfnamefont {Warren}\ \bibnamefont {Weckesser}}, \bibinfo {author} {\bibfnamefont {Jonathan}\ \bibnamefont {Bright}}, \bibinfo {author} {\bibfnamefont {St{\'e}fan~J.}\ \bibnamefont {{van der Walt}}}, \bibinfo {author} {\bibfnamefont {Matthew}\ \bibnamefont {Brett}}, \bibinfo {author} {\bibfnamefont {Joshua}\ \bibnamefont {Wilson}}, \bibinfo {author} {\bibfnamefont {K.~Jarrod}\ \bibnamefont {Millman}}, \bibinfo {author}
  {\bibfnamefont {Nikolay}\ \bibnamefont {Mayorov}}, \bibinfo {author} {\bibfnamefont {Andrew R.~J.}\ \bibnamefont {Nelson}}, \bibinfo {author} {\bibfnamefont {Eric}\ \bibnamefont {Jones}}, \bibinfo {author} {\bibfnamefont {Robert}\ \bibnamefont {Kern}}, \bibinfo {author} {\bibfnamefont {Eric}\ \bibnamefont {Larson}}, \bibinfo {author} {\bibfnamefont {C~J}\ \bibnamefont {Carey}}, \bibinfo {author} {\bibfnamefont {{\.I}lhan}\ \bibnamefont {Polat}}, \bibinfo {author} {\bibfnamefont {Yu}~\bibnamefont {Feng}}, \bibinfo {author} {\bibfnamefont {Eric~W.}\ \bibnamefont {Moore}}, \bibinfo {author} {\bibfnamefont {Jake}\ \bibnamefont {{VanderPlas}}}, \bibinfo {author} {\bibfnamefont {Denis}\ \bibnamefont {Laxalde}}, \bibinfo {author} {\bibfnamefont {Josef}\ \bibnamefont {Perktold}}, \bibinfo {author} {\bibfnamefont {Robert}\ \bibnamefont {Cimrman}}, \bibinfo {author} {\bibfnamefont {Ian}\ \bibnamefont {Henriksen}}, \bibinfo {author} {\bibfnamefont {E.~A.}\ \bibnamefont {Quintero}}, \bibinfo {author} {\bibfnamefont
  {Charles~R.}\ \bibnamefont {Harris}}, \bibinfo {author} {\bibfnamefont {Anne~M.}\ \bibnamefont {Archibald}}, \bibinfo {author} {\bibfnamefont {Ant{\^o}nio~H.}\ \bibnamefont {Ribeiro}}, \bibinfo {author} {\bibfnamefont {Fabian}\ \bibnamefont {Pedregosa}}, \bibinfo {author} {\bibfnamefont {Paul}\ \bibnamefont {{van Mulbregt}}}, \ and\ \bibinfo {author} {\bibnamefont {{SciPy 1.0 Contributors}}},\ }\bibfield  {title} {\enquote {\bibinfo {title} {{{SciPy} 1.0: Fundamental Algorithms for Scientific Computing in Python}},}\ }\href {\doibase 10.1038/s41592-019-0686-2} {\bibfield  {journal} {\bibinfo  {journal} {Nature Methods}\ }\textbf {\bibinfo {volume} {17}},\ \bibinfo {pages} {261--272} (\bibinfo {year} {2020})}\BibitemShut {NoStop}%
\bibitem [{\citenamefont {Graffitti}\ \emph {et~al.}(2018)\citenamefont {Graffitti}, \citenamefont {Barrow}, \citenamefont {Proietti}, \citenamefont {Kundys},\ and\ \citenamefont {Fedrizzi}}]{graffitti2018independent}%
  \BibitemOpen
  \bibfield  {author} {\bibinfo {author} {\bibfnamefont {Francesco}\ \bibnamefont {Graffitti}}, \bibinfo {author} {\bibfnamefont {Peter}\ \bibnamefont {Barrow}}, \bibinfo {author} {\bibfnamefont {Massimiliano}\ \bibnamefont {Proietti}}, \bibinfo {author} {\bibfnamefont {Dmytro}\ \bibnamefont {Kundys}}, \ and\ \bibinfo {author} {\bibfnamefont {Alessandro}\ \bibnamefont {Fedrizzi}},\ }\bibfield  {title} {\enquote {\bibinfo {title} {Independent high-purity photons created in domain-engineered crystals},}\ }\href@noop {} {\bibfield  {journal} {\bibinfo  {journal} {Optica}\ }\textbf {\bibinfo {volume} {5}},\ \bibinfo {pages} {514--517} (\bibinfo {year} {2018})}\BibitemShut {NoStop}%
\bibitem [{\citenamefont {Pickston}\ \emph {et~al.}(2021)\citenamefont {Pickston}, \citenamefont {Graffitti}, \citenamefont {Barrow}, \citenamefont {Morrison}, \citenamefont {Ho}, \citenamefont {Bra\'{n}czyk},\ and\ \citenamefont {Fedrizzi}}]{Pickston:21}%
  \BibitemOpen
  \bibfield  {author} {\bibinfo {author} {\bibfnamefont {Alexander}\ \bibnamefont {Pickston}}, \bibinfo {author} {\bibfnamefont {Francesco}\ \bibnamefont {Graffitti}}, \bibinfo {author} {\bibfnamefont {Peter}\ \bibnamefont {Barrow}}, \bibinfo {author} {\bibfnamefont {Christopher~L.}\ \bibnamefont {Morrison}}, \bibinfo {author} {\bibfnamefont {Joseph}\ \bibnamefont {Ho}}, \bibinfo {author} {\bibfnamefont {Agata~M.}\ \bibnamefont {Bra\'{n}czyk}}, \ and\ \bibinfo {author} {\bibfnamefont {Alessandro}\ \bibnamefont {Fedrizzi}},\ }\bibfield  {title} {\enquote {\bibinfo {title} {Optimised domain-engineered crystals for pure telecom photon sources},}\ }\href {\doibase 10.1364/OE.416843} {\bibfield  {journal} {\bibinfo  {journal} {Opt. Express}\ }\textbf {\bibinfo {volume} {29}},\ \bibinfo {pages} {6991--7002} (\bibinfo {year} {2021})}\BibitemShut {NoStop}%
\bibitem [{\citenamefont {Schr{\"o}dinger}(1935)}]{schrodinger1935discussion}%
  \BibitemOpen
  \bibfield  {author} {\bibinfo {author} {\bibfnamefont {Erwin}\ \bibnamefont {Schr{\"o}dinger}},\ }\bibfield  {title} {\enquote {\bibinfo {title} {Discussion of probability relations between separated systems},}\ }in\ \href@noop {} {\emph {\bibinfo {booktitle} {Mathematical Proceedings of the Cambridge Philosophical Society}}},\ Vol.~\bibinfo {volume} {31}\ (\bibinfo {organization} {Cambridge University Press},\ \bibinfo {year} {1935})\ pp.\ \bibinfo {pages} {555--563}\BibitemShut {NoStop}%
\bibitem [{\citenamefont {Wiseman}\ \emph {et~al.}(2007)\citenamefont {Wiseman}, \citenamefont {Jones},\ and\ \citenamefont {Doherty}}]{Wiseman}%
  \BibitemOpen
  \bibfield  {author} {\bibinfo {author} {\bibfnamefont {H.~M.}\ \bibnamefont {Wiseman}}, \bibinfo {author} {\bibfnamefont {S.~J.}\ \bibnamefont {Jones}}, \ and\ \bibinfo {author} {\bibfnamefont {A.~C.}\ \bibnamefont {Doherty}},\ }\bibfield  {title} {\enquote {\bibinfo {title} {Steering, entanglement, nonlocality, and the einstein-podolsky-rosen paradox},}\ }\href {\doibase 10.1103/PhysRevLett.98.140402} {\bibfield  {journal} {\bibinfo  {journal} {Phys. Rev. Lett.}\ }\textbf {\bibinfo {volume} {98}},\ \bibinfo {pages} {140402} (\bibinfo {year} {2007})}\BibitemShut {NoStop}%
\bibitem [{\citenamefont {Proietti}\ \emph {et~al.}(2021)\citenamefont {Proietti}, \citenamefont {Ho}, \citenamefont {Grasselli}, \citenamefont {Barrow}, \citenamefont {Malik},\ and\ \citenamefont {Fedrizzi}}]{proietti2021NQKD}%
  \BibitemOpen
  \bibfield  {author} {\bibinfo {author} {\bibfnamefont {Massimiliano}\ \bibnamefont {Proietti}}, \bibinfo {author} {\bibfnamefont {Joseph}\ \bibnamefont {Ho}}, \bibinfo {author} {\bibfnamefont {Federico}\ \bibnamefont {Grasselli}}, \bibinfo {author} {\bibfnamefont {Peter}\ \bibnamefont {Barrow}}, \bibinfo {author} {\bibfnamefont {Mehul}\ \bibnamefont {Malik}}, \ and\ \bibinfo {author} {\bibfnamefont {Alessandro}\ \bibnamefont {Fedrizzi}},\ }\bibfield  {title} {\enquote {\bibinfo {title} {Experimental quantum conference key agreement},}\ }\href {\doibase 10.1126/sciadv.abe0395} {\bibfield  {journal} {\bibinfo  {journal} {Science Advances}\ }\textbf {\bibinfo {volume} {7}},\ \bibinfo {pages} {eabe0395} (\bibinfo {year} {2021})}\BibitemShut {NoStop}%
\bibitem [{\citenamefont {Khan}\ \emph {et~al.}(2018)\citenamefont {Khan}, \citenamefont {Solmeyer}, \citenamefont {Balu},\ and\ \citenamefont {Humble}}]{khan2018quantum}%
  \BibitemOpen
  \bibfield  {author} {\bibinfo {author} {\bibfnamefont {Faisal~Shah}\ \bibnamefont {Khan}}, \bibinfo {author} {\bibfnamefont {Neal}\ \bibnamefont {Solmeyer}}, \bibinfo {author} {\bibfnamefont {Radhakrishnan}\ \bibnamefont {Balu}}, \ and\ \bibinfo {author} {\bibfnamefont {Travis~S}\ \bibnamefont {Humble}},\ }\bibfield  {title} {\enquote {\bibinfo {title} {Quantum games: a review of the history, current state, and interpretation},}\ }\href {\doibase 10.1007/s11128-018-2082-8} {\bibfield  {journal} {\bibinfo  {journal} {Quantum Information Processing}\ }\textbf {\bibinfo {volume} {17}},\ \bibinfo {pages} {1--42} (\bibinfo {year} {2018})}\BibitemShut {NoStop}%
\end{thebibliography}%

\end{document}